\documentclass[twocolumn,trackchanges]{aastex62}

\graphicspath{{./}{figures/}}

\received{XXXX}
\revised{XXXX}
\accepted{XXXX}
\submitjournal{AJ}
\shorttitle{Star formation in WLM}
\shortauthors{Mondal et al.}

\begin{document}

\title{UVIT imaging of WLM : Demographics of star forming regions in the nearby dwarf irregular galaxy}

\author{Chayan Mondal}
\affiliation{Indian Institute of Astrophysics, Koramangala II Block, Bangalore-560034, India}
\email{chayan@iiap.res.in, mondalchayan1991@gmail.com}

\author{Annapurni Subramaniam}
\affiliation{Indian Institute of Astrophysics, Koramangala II Block, Bangalore-560034, India}

\author{Koshy George}
\affiliation{Indian Institute of Astrophysics, Koramangala II Block, Bangalore-560034, India}

\begin{abstract}
We present a study of star forming regions and its demographics in the nearby dwarf irregular galaxy WLM using the Ultra-Violet Imaging Telescope (UVIT) multi band observations in three filters F148W, N245M and N263M. We find that the UV emission is extended at least up to 1.7 kpc, with the NUV emission more extended than the FUV. We create UV color maps ((F148W$-$N245M) and (F148W$-$N263M)) to study the temperature morphology of young stellar complexes with the help of theoretical models. We identify several complexes with temperature T $>$ 17500 K which are likely to be the OB associations present in the galaxy. These complexes show good spatial correlation with the H$\alpha$ emitting regions, H$~$I distribution and HST detected hot stars. The hot star forming regions are found to be clumpy in nature and show a hierarchical structure, with sizes in the range of 4 - 50 pc, with a large number with sizes $<$ 10 pc. The south western part of the galaxy shows many hot star forming regions, high level of H$\alpha$ emission and low column density of H$~$I which altogether denote a vigorous recent star formation. WLM is likely to have a large fraction of low mass compact star forming regions with mass M $< 10^3 M_{\odot}$, in agreement with the size and mass of the CO clouds. We estimate the star formation rate of WLM to be $\sim$ 0.008 $M_{\odot}/yr$, which is similar to the average value measured for nearby dwarf irregular galaxies.
\end{abstract}


\section{Introduction}

Dwarf galaxies are the most abundant type of galaxy in the universe. These are the building blocks of the galaxy formation and evolution, and formed when the universe was very young. In the hierarchical merging LCDM paradigm, the stellar halos of the
massive galaxies in the local universe are thought to have accumulated
from the
continuous merging of dwarf galaxies. Despite their numbers and proximity, they are still among the least understood objects.
Understanding the factors that allow these dwarf galaxies to survive to the present day may explain the
discrepancies between observed and predicted distribution of such galaxies in and around the Milky Way.
This requires an understanding of how internal and environmental effects are expected to shape the evolution of low-mass systems like dwarf galaxies. These galaxies have low metallicity and shallow potential well, and hence understanding how star formation proceeds in such regimes is crucial to study their counterparts in the early universe \citep{weisz2011,bianchi2012}. While the massive galaxies formed most of their stars in the first few Gyr, the dwarf galaxies are
found to form stars over the entire cosmic time, resulting in a wide variety of star formation histories \citep{cignoni2018,tolstoy2009,mcquinn2015} and
specific star formation rates, ranging from nearly inactive dwarfs to very active Blue Compact Dwarfs (BCDs).\\\\
Nearby dwarf galaxies offer a unique opportunity to study star formation in a low mass and metal poor environment. \citet{leaman2012} presented a detailed discussion on the possible mechanisms which keep the star formation alive in dwarf galaxies, despite their shallow potential well. The absence of spiral density wave makes them an ideal laboratory to understand the effect of internal triggering mechanisms, such as feedback from massive stars \citep{hunter1997}.  These internal triggers mainly have localised effect in the interstellar medium of the galaxy. In order to understand these internal triggers, it is important to identify the distribution of hot massive stars in the galaxy. As the young and hot stars are easily detected in the ultraviolet (UV), the extent and structure of UV disk for any dwarf galaxy is an excellent probe to study the global star forming activities in the galaxy \citep{hunter2010}.\\\\

WLM is a faint dwarf irregular galaxy of the Local group, located at a distance of 995 kpc \citep{urbaneja2008}. It was first discovered by Wolf (1909) and later confirmed by Lundmark and Melotte (1926) and thus got the name as WLM (Wolf-Lundmark-Melotte) \citep{dolphin2000}. The nearest neighbour of WLM, located 175 kpc away, is a small dSph galaxy Cetus \citep{whiting1999}, whereas the massive spirals (MW, M31, M33) are about 1 Mpc away \citep{minniti1997}. Due to its isolated location in the sky, it is believed that the galaxy has not interacted with other galaxies since its creation. This is demonstrated in the figure 1 of \citet{leaman2012}. A study on Leo A, an isolated dwarf irregular galaxy, performed by \citet{cole2007} showed that the star formation pattern and chemical evolution can be different in isolated dwarfs than those present in a galaxy group \citep{weisz2011}. 
The metal poor environment ($logZ = -0.87$, \citet{urbaneja2008}) of WLM has made it an ideal sample to study such isolated systems. Parameters of WLM are presented in Table \ref{wlm}.\\\\

Young massive OB stellar populations are found to have intense UV continuum flux due to their high temperature. Imaging studies in UV broad bands can identify the young and massive groups of stars, whereas optical bands get contribution from the low mass stars. The UV images of nearby star forming galaxies from GALEX mission made a phenomenal contribution in the last decade \citep{gildepaz2007,kara2013,thilker2007}. The Ultra-violet imaging telescope (UVIT) \citep{kumar2012} telescope on ASTROSAT, with its superior spatial resolution is expected to provide better insight into the physics of star formation in galaxies \citep{koshy2018a,koshy2018b,koshy2018c}. Here, we present the details of star formation in the WLM galaxy, based on the UVIT observations in one FUV broad band and two NUV medium band filters.\\\\

Previous photometric studies find this galaxy to have a relatively metal rich younger disk with an old metal poor halo \citep{ferraro1989,minniti1997,dolphin2000,rejkuba2000}. \citet{melena2009} studied the UV bright regions in 11 dwarf irregular galaxies including WLM. They detected 165 bright UV regions of the galaxy using GALEX data and estimated their age and mass, assuming them to be star clusters. The cluster formation was found to show a constant rate which stopped 22 Myr ago. They found the UV disk to be extended more than the H$\alpha$ disk of the galaxy. A recent HST study by \citet{bianchi2012} covered the galaxy in three HST pointings, and detected a rich population of young and hot stars in the 1 - 10 Myr age range.\\\\

H$~$I disk of galaxy is also well studied and found to be extended than the optical or UV disk. \citet{kepley2007} used VLA observation to map the H$~$I density distribution of the galaxy and found a hook like structure near the center. They concluded that the star formation propagating out from the center of the galaxy created this pattern. Using ALMA observations \citet{rubio2015} identified several CO cloud cores in the galaxy. The presence of these low mass clouds were a new insight in such a metal poor environment, but revealed the possibility of low mass cluster formation in the galaxy.\\\\

In this study we examined the FUV and NUV emission profile of the WLM, with the help of UVIT multi band observations.  We make use of the broad band FUV filter, F148W, which is similar to the GALEX FUV filter to map the FUV emission. We use two medium band NUV filters N245M and N263M, to study the NUV emission. These filters along with the FUV filter are used to make the UV color maps. One of the filters, N245M is located in the wing of the 217.5nm extinction feature whereas N263M is unaffected, which is used to address the issue of dust extinction. Here we use  the FUV$-$NUV color to derive the temperature of the hot SF regions and estimate their temperature distribution with the help of theoretical models. This study thus provides the demographics of massive OB star in WLM. 
The paper is arranged as follows : The data are presented in section 2, models are discussed in section 3, effect of extinction and metallicity in section 4, analysis in section 5, followed by discussion and summary in sections 6 and 7 respectively.

\begin{table}
\centering
\caption{Properties of WLM}
 \label{wlm}
\resizebox{90mm}{!}{

\begin{tabular}{ccc}
\hline
 Property & Value & Reference\\\hline
 RA & 00:01:57.8 & \citet{gallouet1975}\\
 DEC & -15:27:51.0 & \citet{gallouet1975}\\
 Morphological type & IB(s)m & \citet{devaucouleurs1991}\\
 Distance & 995 $\pm$ 46 kpc & \citet{urbaneja2008}\\
 Metallicity ($log(Z/Z_{\odot})$) & $-0.87 \pm 0.06$ & \citet{urbaneja2008}\\
 Inclination & $69^\circ$ & \cite{devaucouleurs1991}\\
 PA of major axis & $181^\circ$ & \citet{devaucouleurs1991}\\\hline

\end{tabular}
}
\end{table}

\section{Data}
\begin{figure*}
\centering
\includegraphics[width = 6.5in]{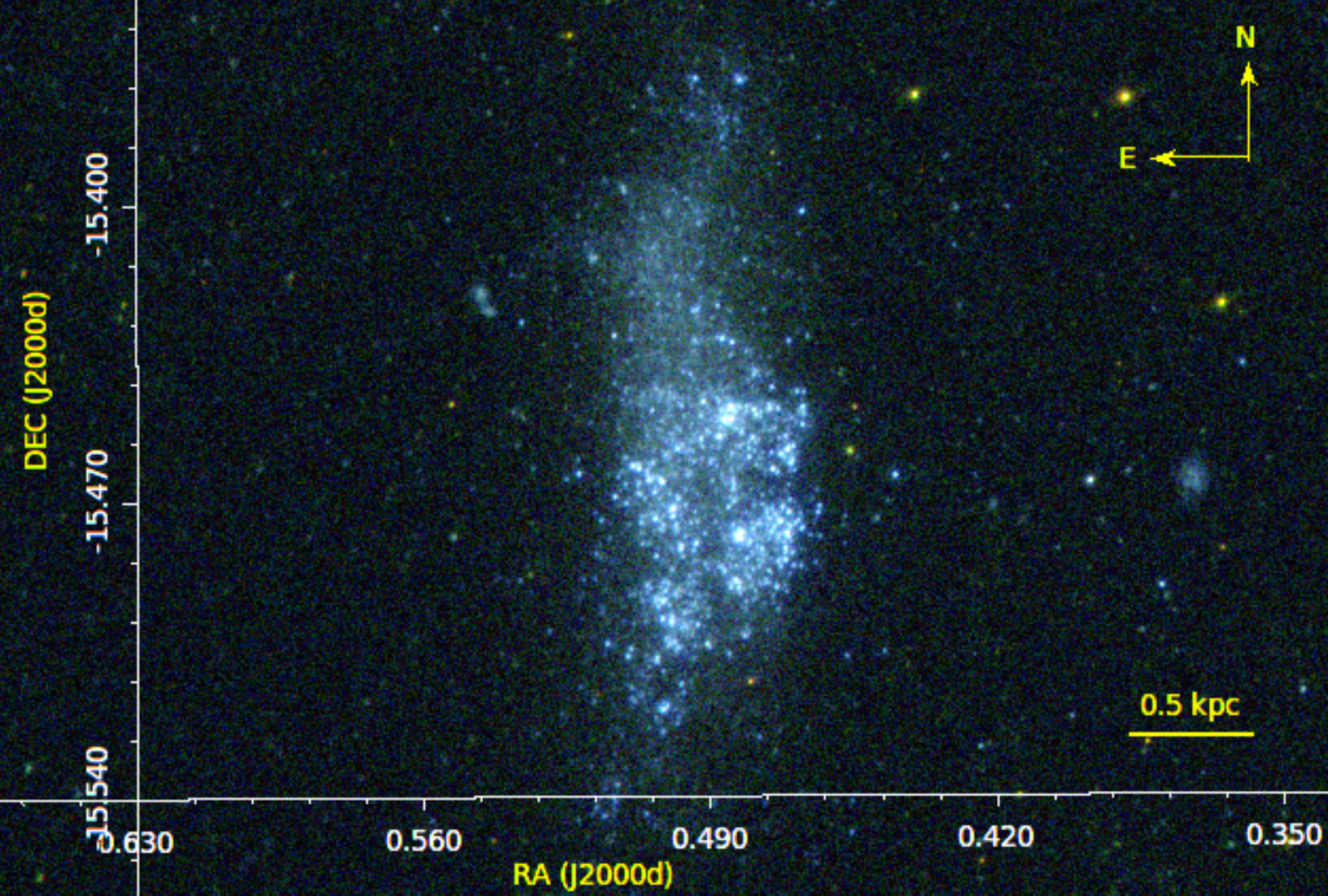}
\caption{False color composite image of the galaxy WLM. The galaxy is observed in three different UVIT filters F148W, N245M and N263M which are represented by blue, green and red colors respectively.}
\label{images_wlm}
\end{figure*}

The UV images of WLM were obtained from UVIT onboard ASTROSAT satellite \citep{kumar2012}. The instrument consists of two co-aligned telescopes, each having an aperture of 375 mm. One of the telescopes observes in FUV band (1300-1800 $\AA{}$) while the other one observes in both NUV (2000-3000 $\AA{}$) and Visible. The telescopes are capable of observing simultaneously in FUV, NUV and visible bands. The visible channel is used to correct the drift pattern in the UV imaging data introduced by the satellite during the course of observation. Each of the FUV, NUV and Visible channel consists of multiple filters having different bandwidths, which indeed is unique, particularly for the UV passbands. The telescope has a 28 arcmin circular field of view with an angular resolution of $\sim$ 1.5 arcsec. It can resolve sources much better than GALEX (4.5 - 5.5 arcsec). The in-orbit calibration of the instrument is completed using HZ4 as calibration source. We refer to \citet{tandon2017} for all the calibration measurements adopted in this paper. The observations for the galaxy WLM were taken in three filters, one in the FUV band (F148W : 1250-1750 $\AA{}$) and two in the NUV bands (N245M : 2148-2710 $\AA{}$ \& N263M : 2461-2845 $\AA{}$). The complete observation was carried out using multiple orbits of the ASTROSAT satellite with a 90 minute period on an equatorial orbit. Data for each of the orbit were corrected for spacecraft drift using the specialized software CCDLAB \citep{postma2017}. We also corrected the data for fixed pattern noise and spatial distortion which are intrinsic to the detector \citep{postma2011,girish2017}. All the images of respective filters are then flat-fielded, co-aligned and combined to produce deeper images which are shown in Figure \ref{images_wlm}. The final exposure time obtained for the filters F148W, N245M and N263M as well as the filter characteristics are tabulated in Table \ref{uvit_obs}. The UVIT images have slightly less than twice the GALEX exposure times of WLM \citep{melena2009}. All images have 4096$\times$4096 pixel dimension with 1 pixel corresponding to 0.4125 arcsec which is equivalent to $\sim$ 2 pc at the distance of WLM. Since there are not enough FUV bright field stars available in the UVIT field of view (Figure \ref{images_wlm}), we used the model PSF image for F148W filter created for the galactic globular cluster NGC 1851 \citep{subramaniam2017}. In case of two NUV filters the model psf were built from the field stars present in the UVIT field. The FWHM of the model psf for both F148W and N263M filter, which are primarily used in our analysis, is  1.5$\arcsec$. The image in N245M filter, which is mainly used for comparison, has a psf of 1.7$\arcsec$. \\\\
We have also used H$\alpha$ image of the galaxy observed with CTIO 4.0 meter telescope \citep{massey2007} and neutral hydrogen (H$~$I) map from VLA telescope \citep{kepley2007} in our study. Both the images were obtained from NASA Extra-galactic Database (NED).  

\begin{table*}
\centering
\caption{Details of UVIT observations}
\label{uvit_obs}
\begin{tabular}{p{2cm}p{2cm}p{3.0cm}p{3cm}p{2cm}p{2cm}}
\hline
Filter & Bandpass ($\AA$) & ZP magnitude (AB) & Unit conversion (erg/sec/cm$^2$/$\AA$) & $\triangle \lambda (\AA)$ & Exposure time (sec)\\\hline
F148W & 1250-1750 & 18.016$\pm$0.01 & 3.09$\times10^{-15}$ & 500 & 2634\\\hline
N245M & 2148-2710 & 18.50$\pm$0.07 & 7.25$\times10^{-16}$ & 280 & 1926\\
N263M & 2461-2845 & 18.18$\pm$0.01 & 8.44$\times10^{-16}$ & 275 & 2370\\\hline

\end{tabular}

\end{table*}

\begin{figure}
\gridline{\fig{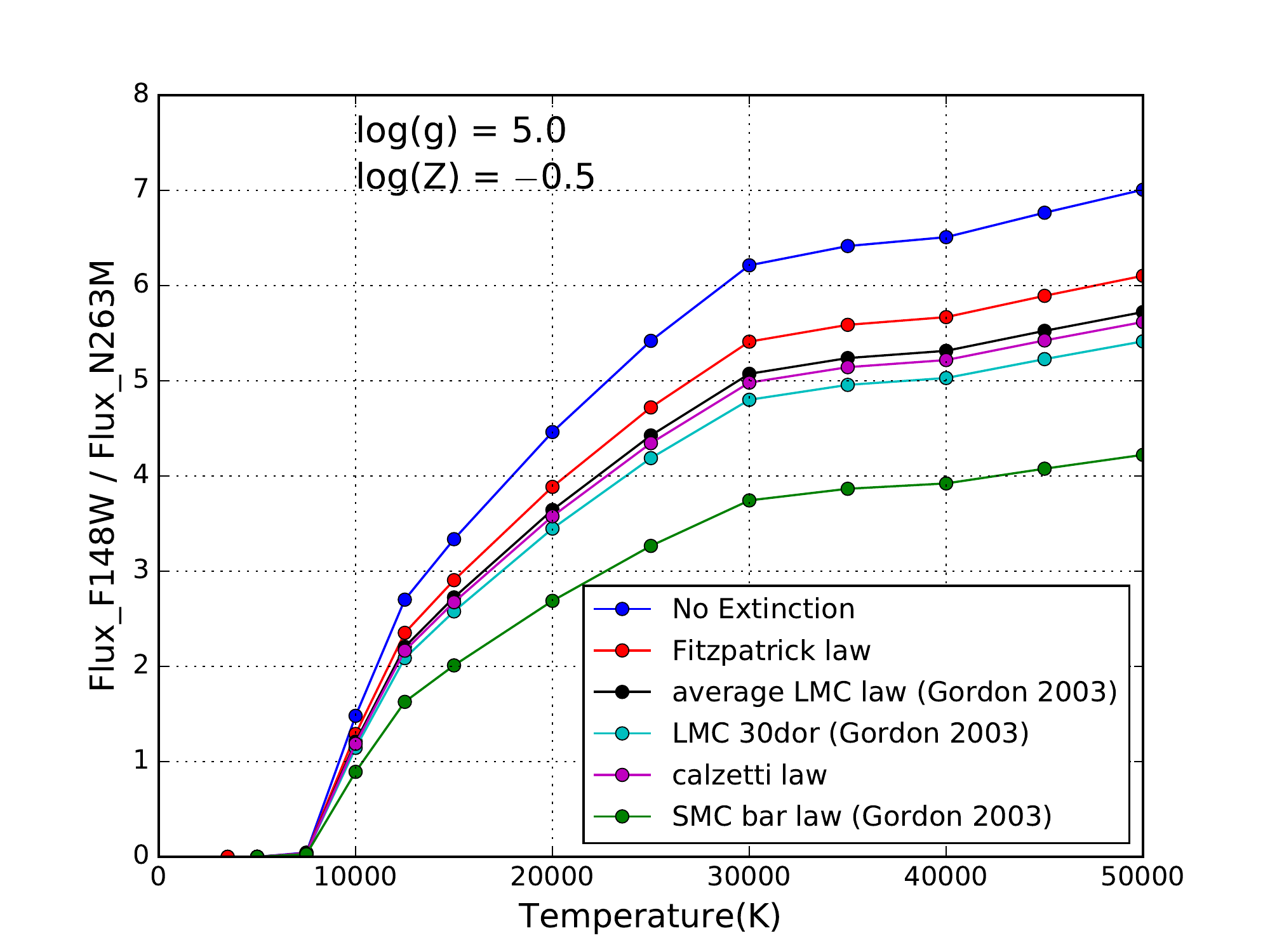}{0.50\textwidth}{(a)}
}
\gridline{\fig{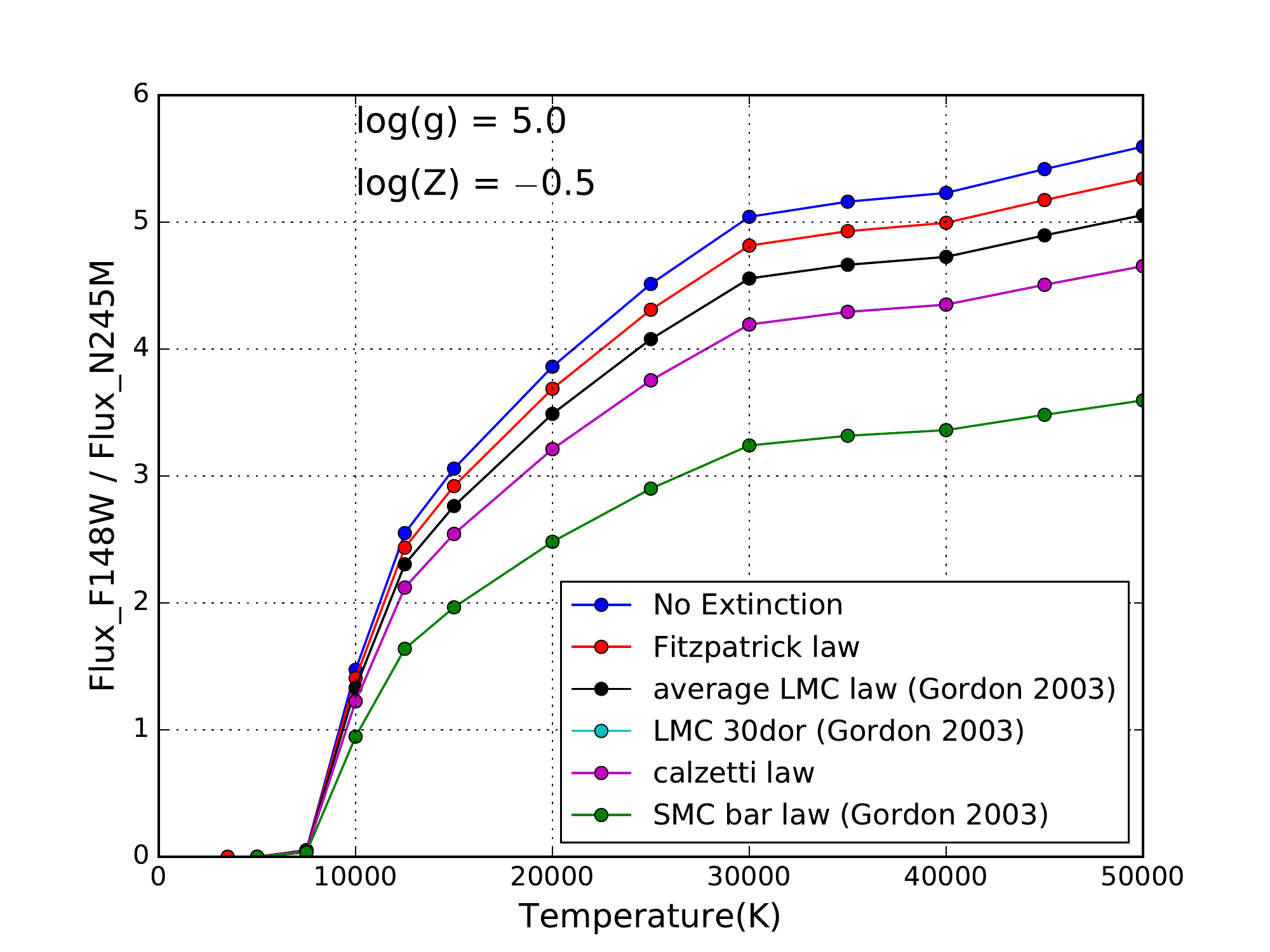}{0.50\textwidth}{(b)}
} 
 \caption{The relation between UV flux ratio and temperature, estimated using Kurucz spectra, for log(g) $=$ 5.0 and log(Z) $= -0.5$ . The F148W/N263M and F148W/N245M flux ratios are shown in Figure (a) and (b) respectively. The blue curves in both the figures are without extinction correction. The other curves are generated by applying extinction on blue curve for different extinction laws mentioned in the figure.}
 \label{kuruz_temp}
 \end{figure}

\section{Theoretical models}

As we plan to identify hot star forming regions based on the UV colors, we need 
to establish a relation between the UV colors (i.e., the flux ratio between UVIT filters)
and temperatures. We used  \citet{castelli2004} stellar models that offer stellar spectrum for different temperature between 3500 - 50000 K, for deriving the above relation. For a fixed temperature, individual spectra are available for different surface gravity (log(g)) and metallicity (log(Z)). We fixed the metallicity as log(Z)$=-0.5$ (which is close to that of WLM \citep{urbaneja2008}, Table \ref{wlm}) and surface gravity log(g)$=5.0$ and selected 13 spectra in the temperature range 3,500 - 50,000 K. The spectra are then convolved with the effective area profile of UVIT filters to calculate the expected flux \citep{tandon2017}.

We created diagnostic plots as shown in Figure \ref{kuruz_temp} (blue curve) by taking the ratio of fluxes for two different combinations of filters (i.e. $Flux_{F148W}/Flux_{N263M}$ in Figure (a) and $Flux_{F148W}/Flux_{N245M}$ in (b)). The other curves in the same plots are generated after including the effect of extinction according to different extinction laws, which is discussed in the next section. These curves are used to estimate the temperature distribution as well as in identifying the potential locations of OB associations.\\\\

\begin{figure}
\begin{center}
\includegraphics[width=3.7in]{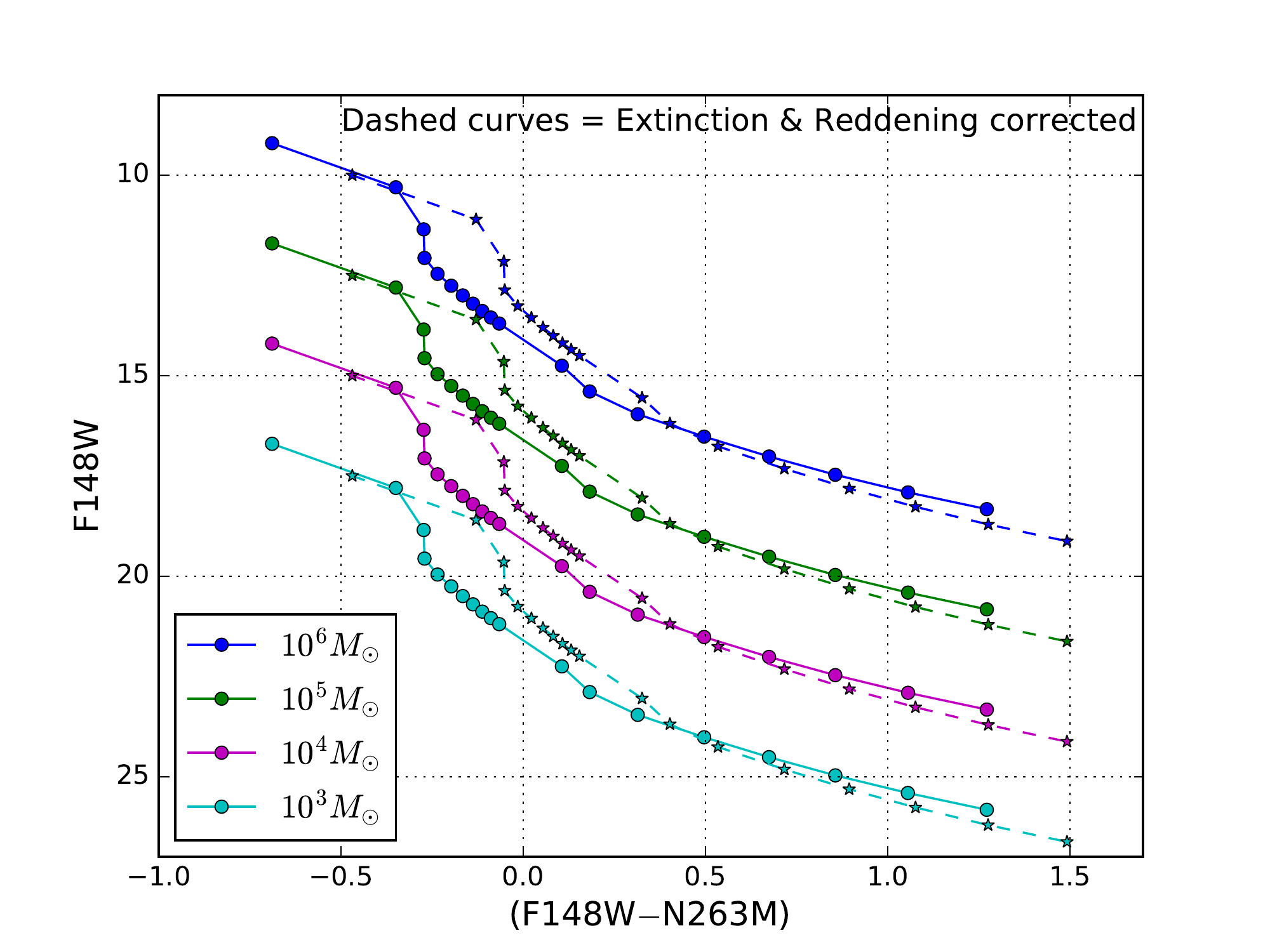} 
 \caption{Starburst99 model generated color-magnitude plots for simple stellar population. The Figure shows F148W magnitude 
    with (F148W$-$N263M) color. Different curves (continuous line) signify four different
    total cluster masses ($10^6 M_\odot$, $10^5 M_\odot$, 
$10^4 M_\odot$, $10^3 M_\odot$). The dashed lines show the extinction and reddening corrected values for each model curve. The points shown in each curve are for different ages starting from 1 Myr to 900 Myr 
(increasing along the color axis) with age interval 10 Myr for 1 to 100 Myr range and 100 Myr for 100 to 900 Myr. The value of other parameters adopted for this figure is listed in Table \ref{starburst999}.}
 \label{cmd_caf2}
 \end{center}
 \end{figure}

We further used starburst99 simple stellar population (SSP) model \citep{leitherer1999} to generate diagnostic diagrams shown in Figure \ref{cmd_caf2} to characterise the identified compact star forming regions. Starburst99 is a spectrophotometric SSP model for actively star forming galaxy. We obtained starburst99 model data for the parameters listed in Table \ref{starburst999}.
We assumed the star formation to be instantaneous for the estimations. IMF value is taken to be 1.3 (for 0.1 $M_{\odot}$ to 0.5 $M_{\odot}$), 2.3 (for 0.5 $M_{\odot}$ to 120 $M_{\odot}$)
(Kroupa IMF)\citep{kroupa2001} with stellar mass limit as $M_{low}=0.1 M_{\odot}$ and $M_{up}=120 M_{\odot}$, 
which are appropriate 
for studying SSP up to a few 100 Myr. 
We produced model grids for F148W magnitude as a function of (F148W$-$N263M) color. 
In order to convert the model generated fluxes to magnitude values, we adopted a distance of 995 kpc \citep{urbaneja2008} 
for WLM. We considered four different values for the total cluster mass ($10^6 M_\odot$, $10^5 M_\odot$, 
$10^4 M_\odot$, $10^3 M_\odot$) and a fixed metallicity of Z=0.004, to generate Figure \ref{cmd_caf2}. The dashed lines are generated by adding the extinction and reddening to each of the actual model curves (continuous lines) shown for different masses. Points shown in each curve signify different ages from 1 to 900 Myr (increasing along the color axis). From Figure \ref{cmd_caf2} we find that, for a fixed age (equivalently color), massive clusters are brighter.

\begin{table}
\centering
 
\caption{Starburst99 model parameters}
\label{starburst999}
\resizebox{90mm}{!}{
\begin{tabular}{cc}
\hline

 Parameter & Value\\\hline
 Star formation & Instantaneous\\
 Stellar IMF & Kroupa (1.3, 2.3)\\
 Stellar mass limit & 0.1, 0.5, 120 $M_{\odot}$\\
 Total cluster mass & $10^3 M_{\odot}$-$10^6 M_{\odot}$\\
 Stellar evolution track & Geneva (high mass loss)\\
 Metallicity & Z=0.004\\
 Age range & 1-900 Myr\\ \hline

\end{tabular}
}

\end{table}

\section{ Effect of Extinction and Metallicity}
\begin{figure}
\begin{center}
\includegraphics[width=3.7in]{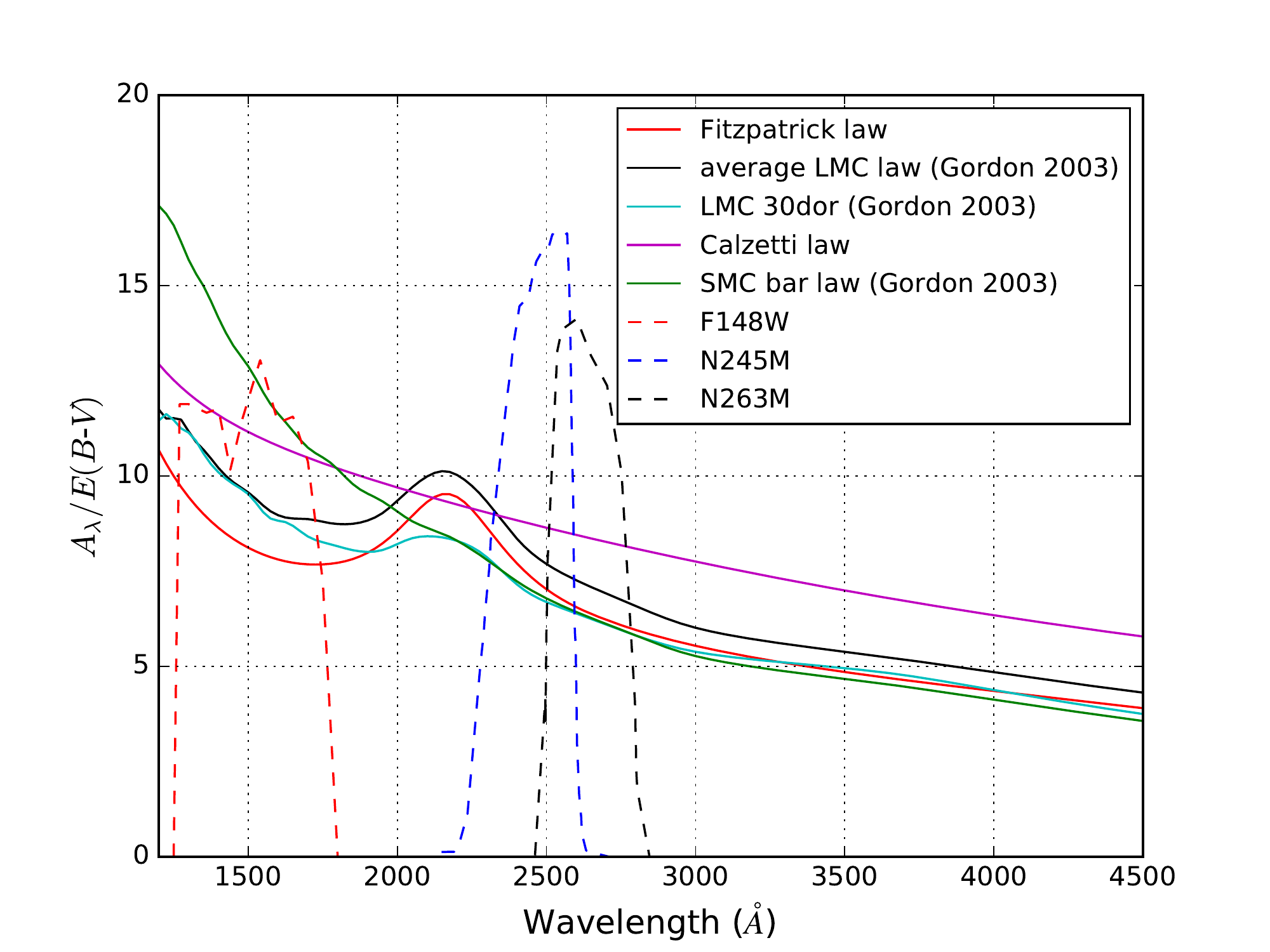} 
 \caption{The variation of extinction coefficient ($R_{\lambda}$, i.e. $A_{\lambda}/E(B-V)$) with wavelength for different extinction laws. The solid lines of different colors represent different laws mentioned in the figure. The scaled effective area curves for three UVIT filters are also shown by the dashed lines.}
 \label{extinction}
 \end{center}
 \end{figure}

\begin{figure}
\begin{center}
\includegraphics[width=3.7in]{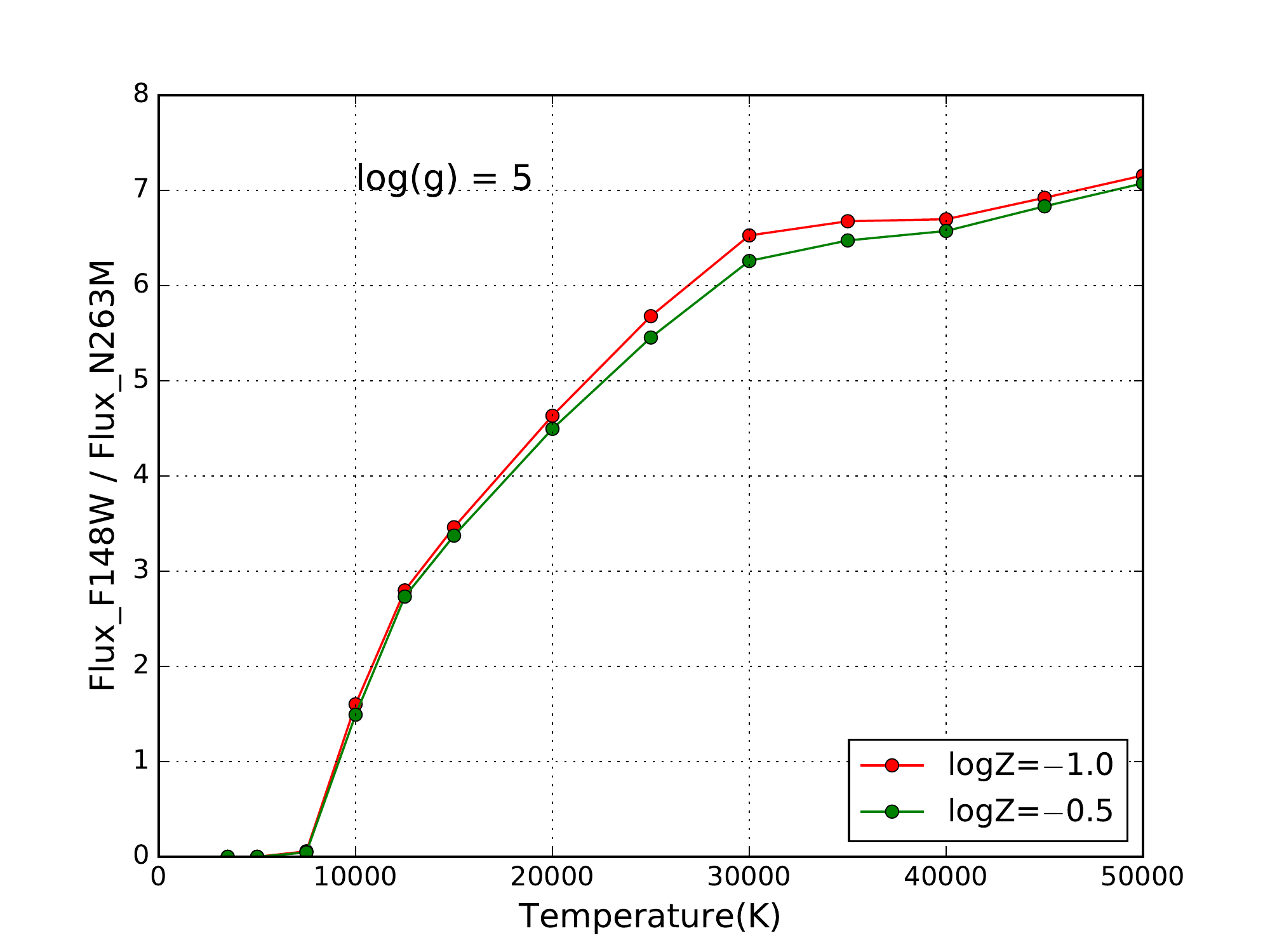} 
 \caption{The flux ratio $Flux_{F148W}/Flux_{N263M}$ is plotted with temperature for two different metallicity log(Z)=$-1.0$ and $-0.5$.}
 \label{kuruz_met}
 \end{center}
 \end{figure}

The behaviour of extinction curve in the UV region shows a noticeable variation for different external galaxies. It is observed that the extinction coefficient in FUV and NUV changes from galaxy to galaxy or even within a galaxy. The 2175 $\AA$ bump is an important feature of the Galactic extinction curve and it also shows characteristic variation for external galaxies. \citet{gordon2003} reported the differential nature of extinction curve in UV for the galaxy LMC, SMC and Milky way. Therefore it is important for us to first understand the nature of extinction law which applies for the star forming regions of WLM and second to understand how the flux in UVIT filters are affected by the extinction law.  In Figure \ref{extinction} we show the behaviour of different extinction laws such as Fitzpatrick law (for Milky Way) \citep{fitzpatrick1999}, Calzetti law (for starburst regions) \citep{calzetti1994}, average LMC, LMC 30 Dor and SMC bar \citep{gordon2003} along with the UVIT filter profiles. The data for the extinction curves are obtained from the extinction calculator \citep{mccall2004} available in NED website. It is quite clear that the N245M filter profile partially overlaps with the 2175 $\AA$ bump of the extinction curve. Assuming the average value of $R_{\lambda}$ within the bandpass of each filter, and adopting E(B$-$V) = 0.082 \citep{gieren2008}, we estimated the extinction value ($A_{\lambda}$) for different extinction laws using the general transformation law given in equation \ref{eq_ext}. The extinction corrected flux ratios after incorporating different extinction laws are shown in Figure \ref{kuruz_temp}. Since the behaviour of extinction in UV for extra-galactic study is highly debated, the simulated plots in Figure \ref{kuruz_temp} highlights the possible uncertainty in the estimated temperature for a fixed flux ratio.\\\\
In their study of the properties of HST identified stars, \citet{bianchi2012} concluded that the most suitable extinction law for the star forming regions of WLM is the average LMC type \citep{gordon2003}. In our study we adopted the UV extinction of WLM to be average LMC type and calculated the value of $R_{\lambda}$ and $A_{\lambda}$ for all three UVIT filters which are given in Table \ref{extinction_table}.\\\\
\begin{table}
\centering
 \caption{Extinction parameters for average LMC law \citep{gordon2003}}
\label{extinction_table}
\resizebox{40mm}{!}{
\begin{tabular}{ccc}
\hline
 Filter & $R_{\lambda}$ & $A_{\lambda}$\\\hline
 F148W & 9.78 & 0.80\\
 N245M & 8.40 & 0.69\\
 N263M & 7.07 & 0.58\\\hline
 \end{tabular}
}
\end{table}
Considering the possibility of a spread in the metallicity across the star forming regions, we explored the effect of metallicity in the estimated temperature. It turns out that as log(Z) changes from $-0.5$ to $-1.0$, the corresponding change in the flux ratio for different temperature is found to be negligible (Figure \ref{kuruz_met}). The metallicity of WLM, measured from a spectroscopic study of blue supergiants present in the galaxy, is reported to be $-0.87$ by \citet{urbaneja2008}. As the effect of metallicity is not very sensitive on the derived temperature, we consider log(Z) $= -0.5$ for the Kurucz model spectra.

 \begin{equation}
A_{\lambda} = R_{\lambda}E(B-V)  
\label{eq_ext}
 \end{equation}
 
\section{Analysis}

\subsection{Extent of UV emission}

\begin{figure}
\begin{center}
\includegraphics[width=2.7in]{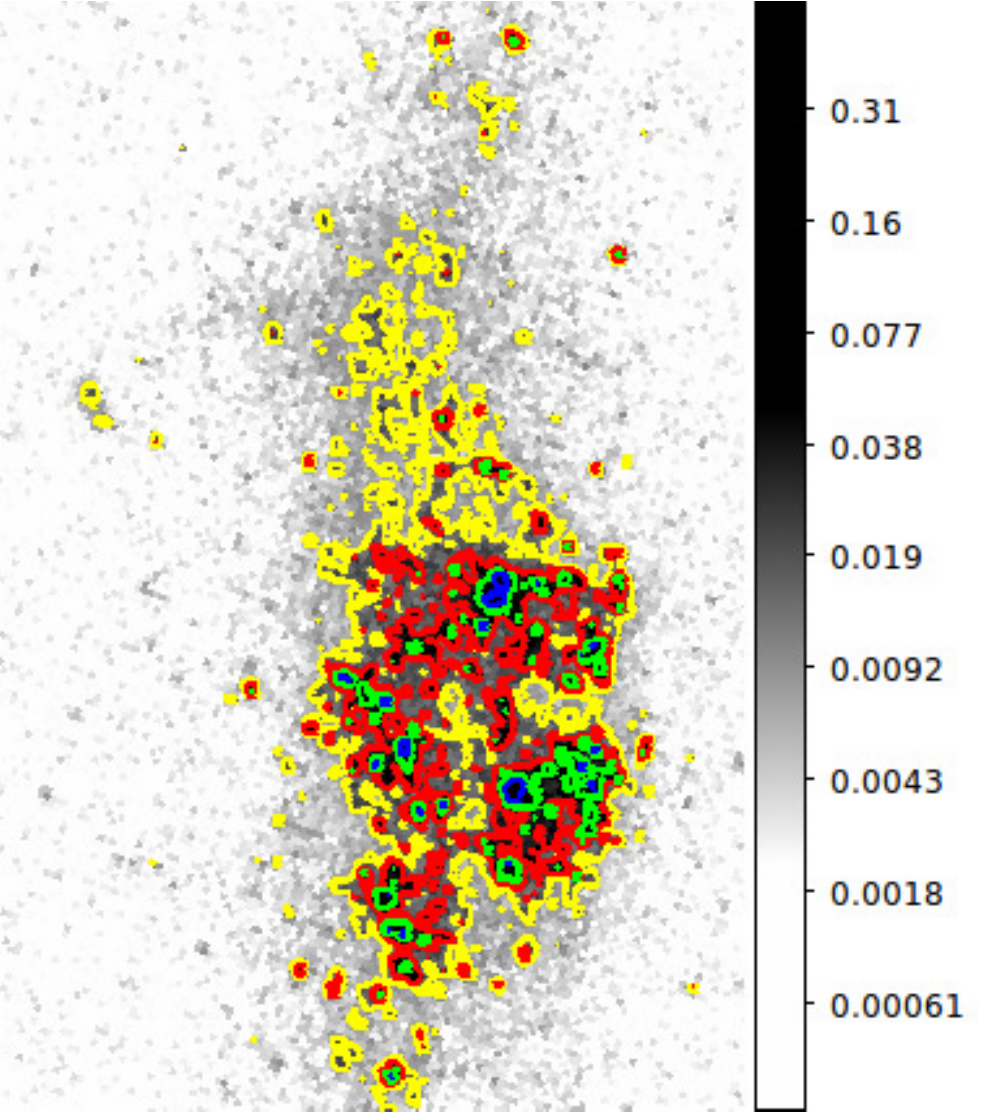} 
 \caption{The background figure is the $512\times512$ F148W band image of WLM with contours plotted for different limits of F148W flux values as mentioned in Table \ref{F148W}. The blue contours represent the brightest (hence most massive) regions of the galaxy. The overall extent of FUV emission of WLM is traced by the yellow contours.}
 \label{caf2_mass}
 \end{center}
 \end{figure}

We consider the F148W band image to study the extent of UV emission in WLM, which in turn traces the expanse of hot and young massive stars. The FUV emitting regions as shown in Figure \ref{images_wlm} are found to be highly structured. Many regions could be massive complexes comprising of a large number of smaller groups, which are not resolved in our images, while some are found to be much smaller, suggestive of smaller groups, and in a few cases individual association of OB stars.
In order to locate the brightest and hence massive complexes, we created a low-resolution map of size 512$\times$512, where each pixel corresponds to 3.3 arcsec which is equivalent to $\sim$ 16 pc at the distance of WLM. 
Each pixel value of the image denotes the integrated count of a stellar group having size $\sim$ 16$\times$16 pc$^2$. As we target to locate large complexes and their sizes, we created contours by fixing the lower and upper limit of flux values which are given in Table \ref{F148W} (shown in Figure \ref{caf2_mass}). The blue contours show pixels brighter than 20 magnitude. These are the most luminous star forming complexes present in the galaxy. We find that at this FUV magnitude, the contributing compact star forming regions are likely to be more massive than 10$^3$ M$_\odot$ (see Figure \ref{cmd_caf2}). The counts per second (CPS) corresponding to this FUV magnitude is 0.16 and there are only a few regions which have CPS more than this value (see figure \ref{caf2_mass}). We are able to identify three luminous (extended) regions which are shown in blue. The green and red contours, which signify relatively less bright regions, are found to be present around these regions, suggestive of a hierarchical structure. The yellow contours generated for pixels fainter than 22 magnitude and brighter than 23 magnitude portray the overall extend of the FUV emission.\\

\begin{figure}
\begin{center}
\includegraphics[width=2.7in]{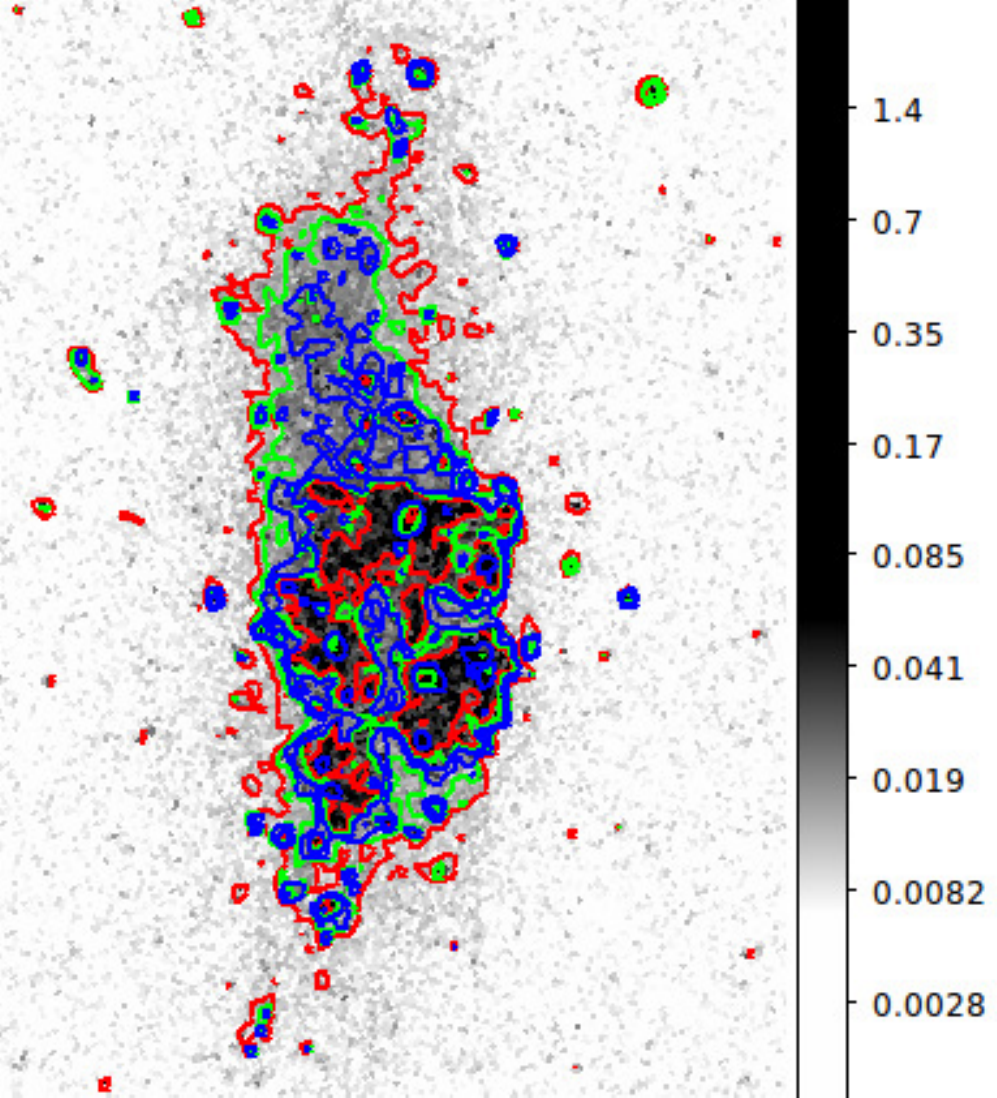} 
 \caption{The background figure is the $512\times512$ N263M band image of WLM. The blue, green and red contours shown in the figure are generated for pixels brighter than 23 magnitude in F148W, N245M and N263M filter images respectively. The FUV emission (blue contours) is seen to be enveloped by both green and red contours representing the NUV emission of WLM.}
 \label{all_mass}
 \end{center}
 \end{figure}

\begin{figure}
\begin{center}
\includegraphics[width=2.7in]{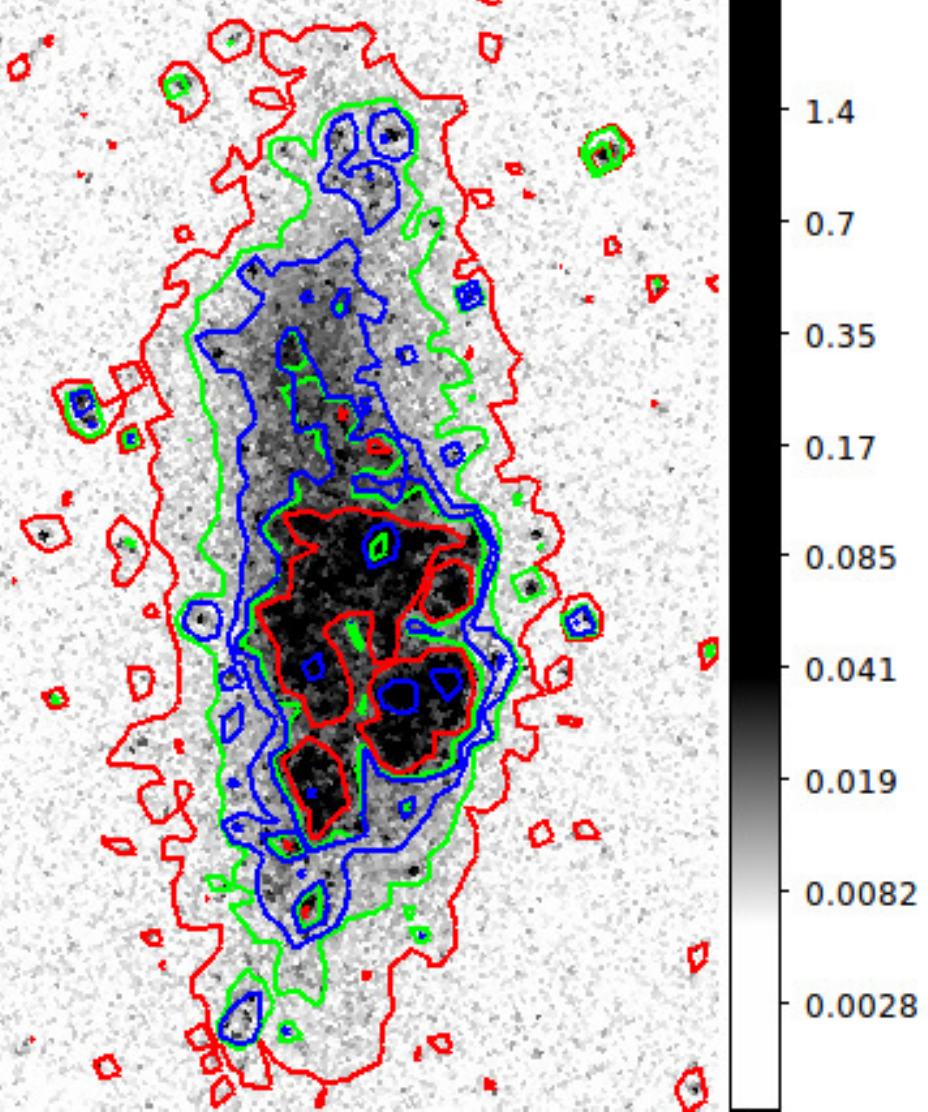} 
 \caption{The background figure is the $512\times512$ N263M band image of WLM. The blue, green and red contours shown in the figure are generated for pixels brighter than 23.80, 23.69 and 23.58 magnitude (adding the effect of extinction) respectively in F148W, N245M and N263M filter images. The extent of NUV emission as traced in N263M filter is found to be more than that of Figure \ref{all_mass}.}
 \label{all_mass_e}
 \end{center}
 \end{figure}

We further studied the structure of emission in two NUV images (N245M and N263M).
Using similar 512$\times$512 images, we generated contours for pixels brighter than 23 magnitude for both the NUV images. For comparison,  
the blue, green and red contours  generated for images in F148W, N245M and N263M band respectively are over laid on the N263M image (Figure \ref{all_mass}). It is clearly seen that the extent of UV emission gradually increases as we go towards longer wavelengths. The red contours, signifying the emission in NUV N263M filter, are found to be stretched more outward than the green (N245M) and blue (F148W) contours. The FUV emission, traced by the blue contours, is found to be present inside the extent of both the NUV emission (green and red contours).  As the contours shown above are likely to be affected by extinction, we apply the extinction correction and reproduce the contours. Applying the average LMC law for extinction \citep{gordon2003}, the extinction corrected magnitudes are 23.80, 23.69 and 23.58 magnitudes, for F148W, N245M and N263M filters, respectively. The blue, green and red contours shown in Figure \ref{all_mass_e} are generated for the above magnitudes in F148W, N245M and N263M filter images respectively. Again the morphology of FUV and NUV emission are found to be similar to Figure \ref{all_mass}. The NUV emission as traced by N263M filter is found to be more extended in Figure \ref{all_mass_e} than in Figure \ref{all_mass}. As the FUV traces hotter and massive population, when compared to the NUV, it is clear that these are located in the inner part of the galaxy, whereas relatively cooler less massive stellar populations are present in the outer regions.
 
 \begin{table}
\centering
 
\caption{Details of flux and magnitudes for contours in the F148W map shown in Figure \ref{caf2_mass}.}
\label{F148W}
\begin{tabular}{p{2.6cm}p{2.5cm}p{1cm}}
\hline

$log(Flux_{F148W})$ range $[erg/sec/cm^2/\AA]$ & Magnitude range (extinction uncorrected) & Contour color\\\hline
$>$ $-$15.3 & $<$ 20 & Blue\\
$-$15.7 to $-$15.3 & $>$ 20 \& $<$ 21 & Green\\
$-$16.1 to $-$15.7 & $>$ 21 \& $<$ 22 & Red\\
$-$16.5 to $-$16.1 & $>$ 22 \& $<$ 23 & Yellow\\\hline
\end{tabular}

\end{table}

\begin{table}
\centering
\caption{Details of flux ratio and temperature for the (F148W$-$N263M) color map, as shown in Figure \ref{caf2_b4}.}
\label{F148W/N263M}

\begin{tabular}{p{3.4cm}p{2.7cm}p{1cm}}
\hline

($Flux_{F148W}/Flux_{N263M}$) range & Temperature range (K) & Contour color\\\hline
$>$5.23 & $>$ 35000 & Blue\\
4.43 to 5.23 & $>$ 25000 \& $<$ 35000 & Red\\
3.18 to 4.43 & $>$ 17500 \& $<$ 25000 & Green\\\hline
\end{tabular}

\end{table}

\subsection{Luminosity density profile}
\label{lum_s}
\begin{figure}
\begin{center}
\includegraphics[width=3.7in]{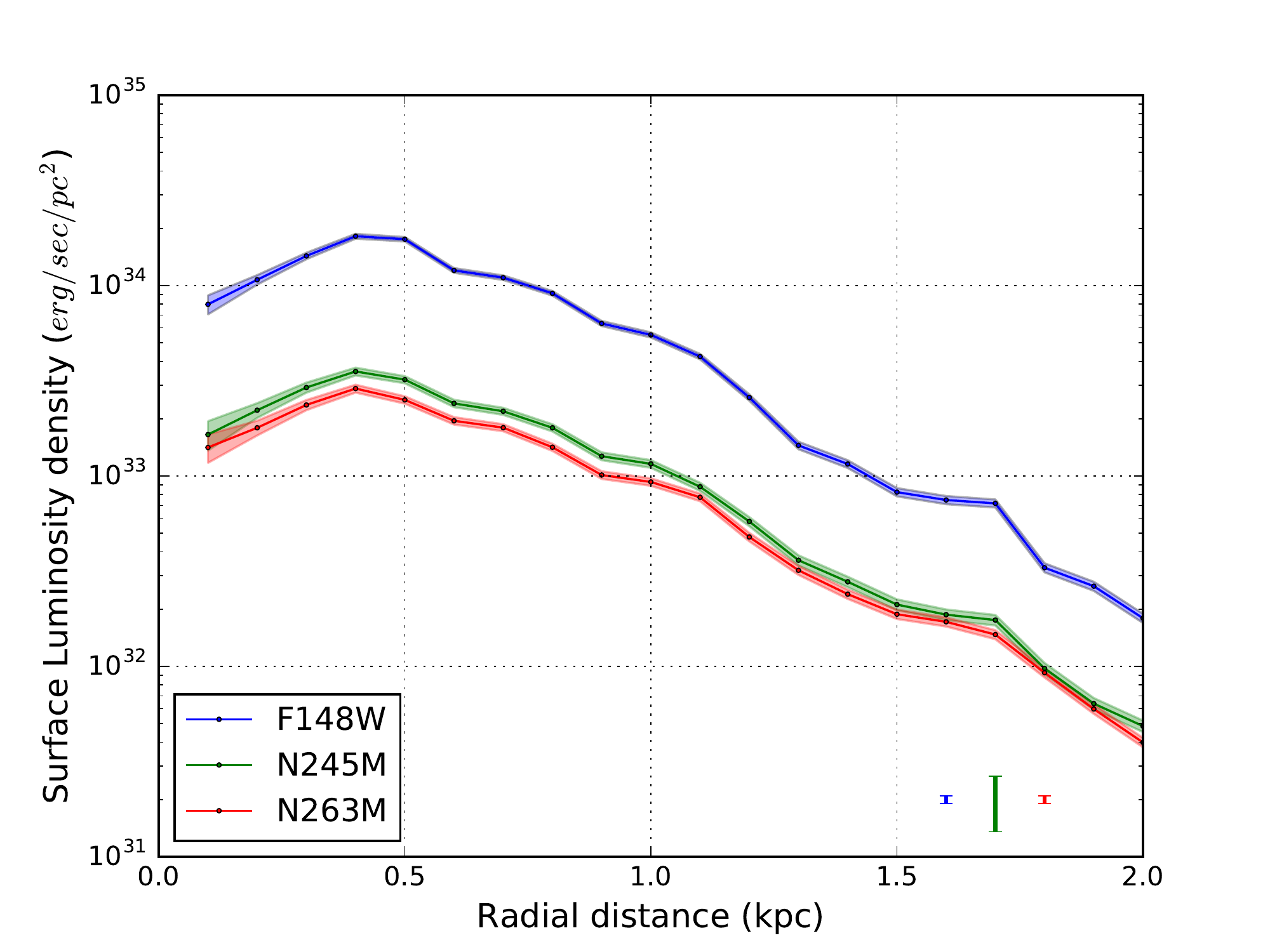} 
 \caption{The profile for surface luminosity density (erg/sec/pc$^2$) with galactocentric distance in kpc for three UVIT filters. The UV emission of WLM is found to be extended at least up to a radius 1.7 kpc. The shaded region corresponding to each curve shows the 5$\sigma$ photometric error whereas the error bars (bottom right) represents 5$\sigma$ error due to the present uncertainty of adopted zero point magnitudes.}
 \label{luminosity}
 \end{center}
 \end{figure}

\begin{figure}

\begin{center}
\includegraphics[width=3.7in]{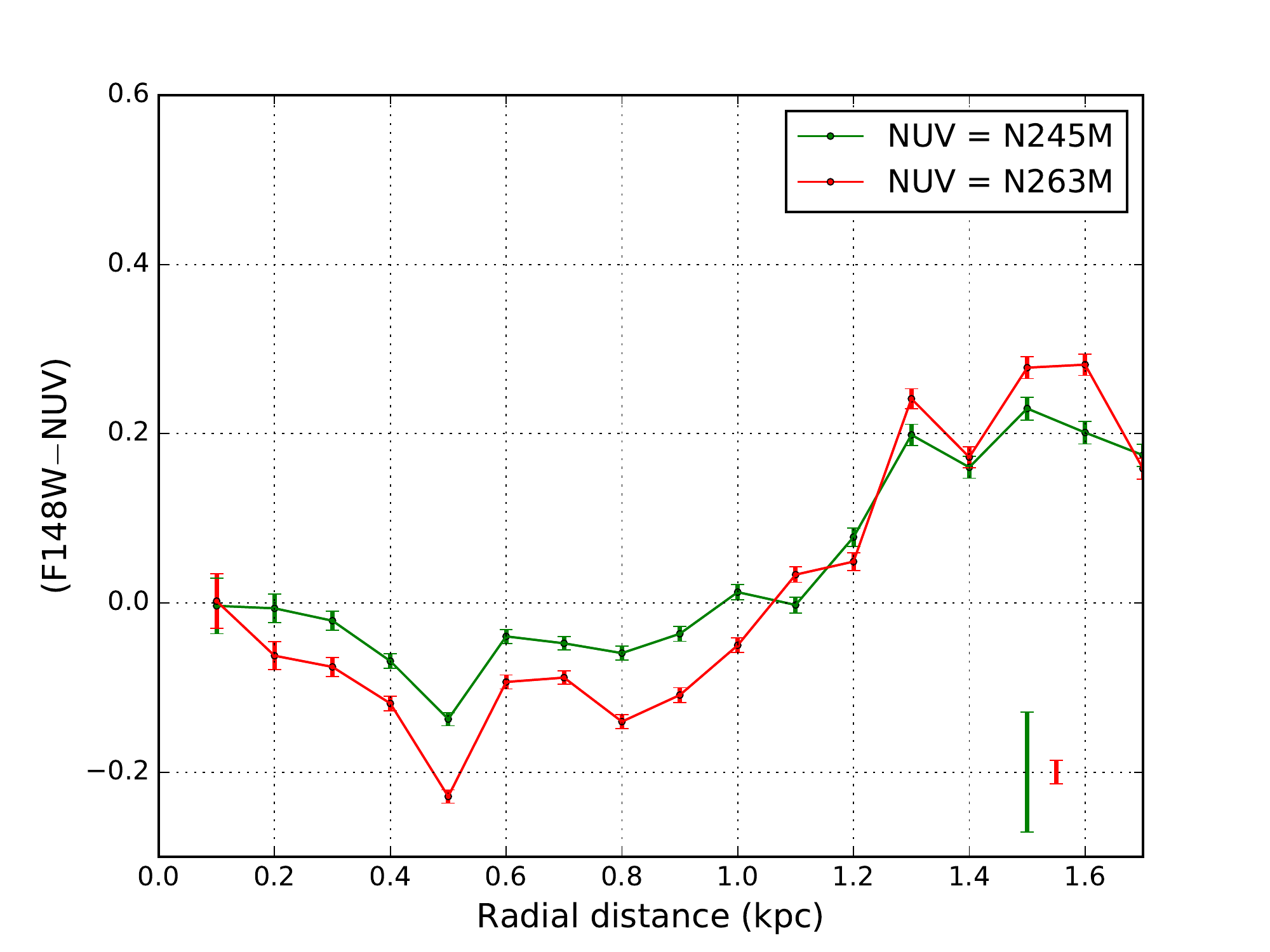} 
 \caption{The variation of (F148W$-$N245M) and (F148W$-$N263M) color is shown with galactocentric distance in kpc. Both the profiles remain nearly flat in the inner 1 kpc whereas an upturn is noticed for the outer part beyond 1 kpc. The photometric errors are shown in the profile and the error bars displayed in the bottom right corner signify the error due to the uncertainty in zero point magnitudes.}
 \label{fn}
 \end{center}
 \end{figure}

The radial variation of luminosity density in UV conveys the characteristic distribution of star forming regions in any active galaxy. Since WLM is observed in one FUV and two NUV filters of UVIT, we constructed the radial surface luminosity density profile in all the filters. Assuming the inclination and position angle of the galaxy (Table \ref{wlm}), we estimated the galactocentric distance  to each pixel in all the three images, using the relation given in \citet{marel2001} . Starting from the galaxy center (RA and Dec given in Table \ref{wlm}), we considered annuli of width 0.1 kpc and measured the total CPS in each individual annuli, up to a radius 2 kpc. The total CPS value is then corrected for background and extinction. The average background in CPS/kpc$^2$ in all the three images are calculated by considering three similar annuli of width 0.1 kpc from radius 2.5 kpc to 2.8 kpc.
The luminosities in each annuli are then estimated by assuming the distance to WLM as 995 kpc, and considering respective value of unit conversion factor and bandwidth of each filter from Table \ref{uvit_obs}. The surface luminosity density (erg/sec/pc$^2$) as a function of radius is shown in Figure \ref{luminosity}. The luminosity density in all three filters shows a dip in the central part of the galaxy and peaks around 0.4 kpc (in F148W it is $\sim$ 0.5 kpc) from the center, which signifies that the star formation is relatively less active near the center of WLM. Beyond 0.4 kpc it decreases again towards the outer part of the galaxy.  
In order to investigate the relative flux variation between the far and near UV, we further plotted (F148W $-$ N245M) and (F148W $-$ N263M)  radial color profile in Figure \ref{fn}. From center to 0.5 kpc, both the colors become bluer with radius. After 0.5 kpc the trend changes and the colors start becoming redder with respect to the value at 0.5 kpc. Beyond 1 kpc the colors become redder, than the average value in the inner 1 kpc. Again (F148W $-$ N263M) color is found to be bluer than (F148W $-$ N245M) in the inner 1 kpc region whereas it becomes almost similar outside 1 kpc. This fact can be interpreted as the outer part beyond 1 kpc has a relatively large extinction and/or less number of massive stars when compared to the inner part.


\subsection{UV color maps}
We use the FUV and NUV images to create UV color maps, to trace the temperature profiles of star forming regions. All the three images are first co-aligned and normalized with respect to their exposure time. The image in F148W filter is divided by the images in N263M and N245M filter to create F148W/N263M and F148W/N245M images respectively. The pixel value of each of these resultant images is actually the CPS ratio in the respective filters.
We infer the temperature corresponding to each pixel of F148W/N263M (or F148W/N245M) image with the help of Figure \ref{kuruz_temp}. We consider the black curve, which is generated according to average LMC type extinction law \citep{gordon2003}, of both Figure \ref{kuruz_temp}a \& b.

 \begin{table}
\centering
\caption{Details of flux ratio and temperature for the (F148W$-$N245M) color map, as shown in Figure \ref{caf2_b13}.}
\label{F148W/N245M}
\begin{tabular}{p{3.4cm}p{2.7cm}p{1cm}}
\hline

($Flux_{F148W}/Flux_{N245M}$) range & Temperature range (K) & Contour color\\\hline
$>$ 4.64 & $>$ 35000 & Blue\\
4.09 to 4.64 & $>$ 25000 \& $<$ 35000 & Red\\
3.15 to 4.09 & $>$ 17500 \& $<$ 25000 & Green\\\hline
\end{tabular}

\end{table}

\subsection{(F148W$-$N263M) color map}
\begin{figure*}
\centering
\gridline{\fig{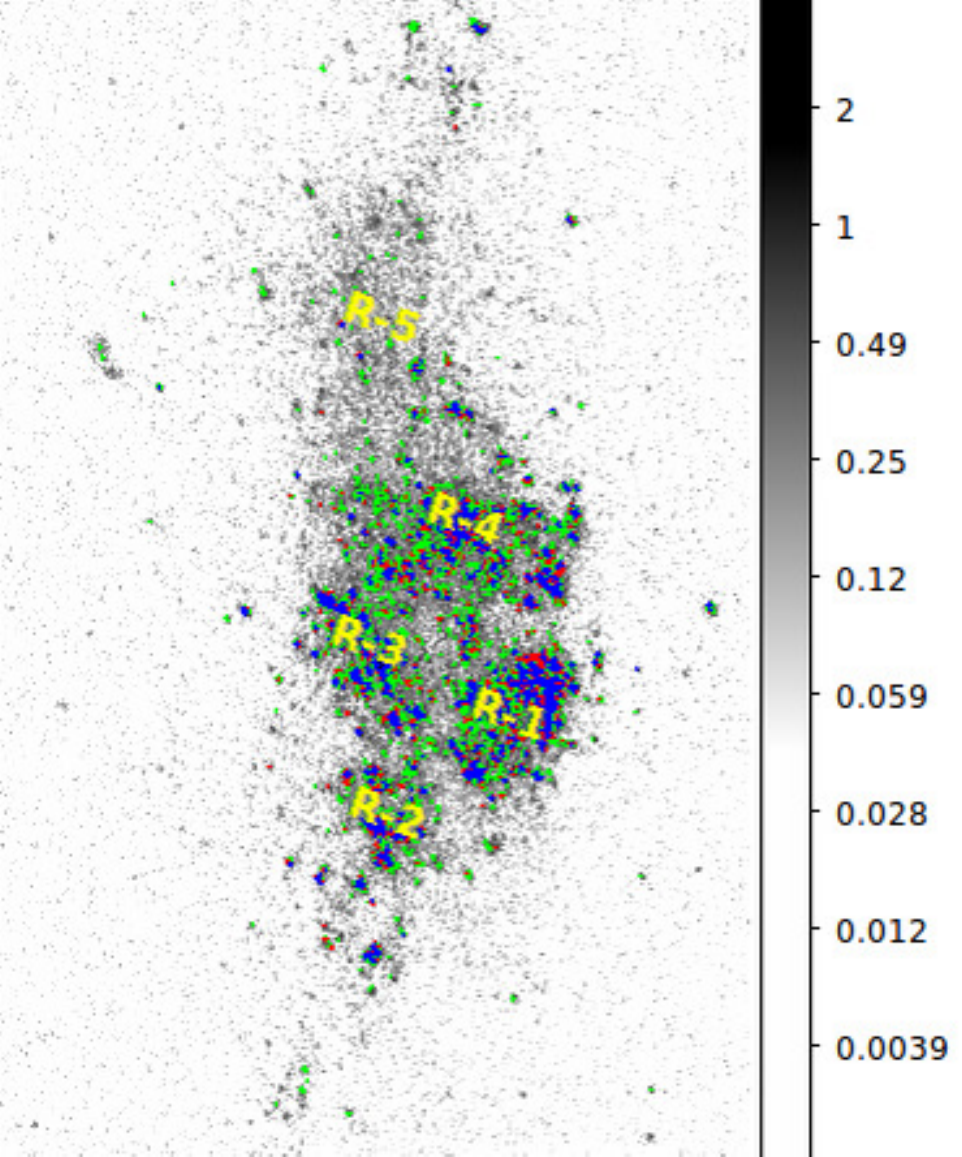}{0.35\textwidth}{(a)}
			\fig{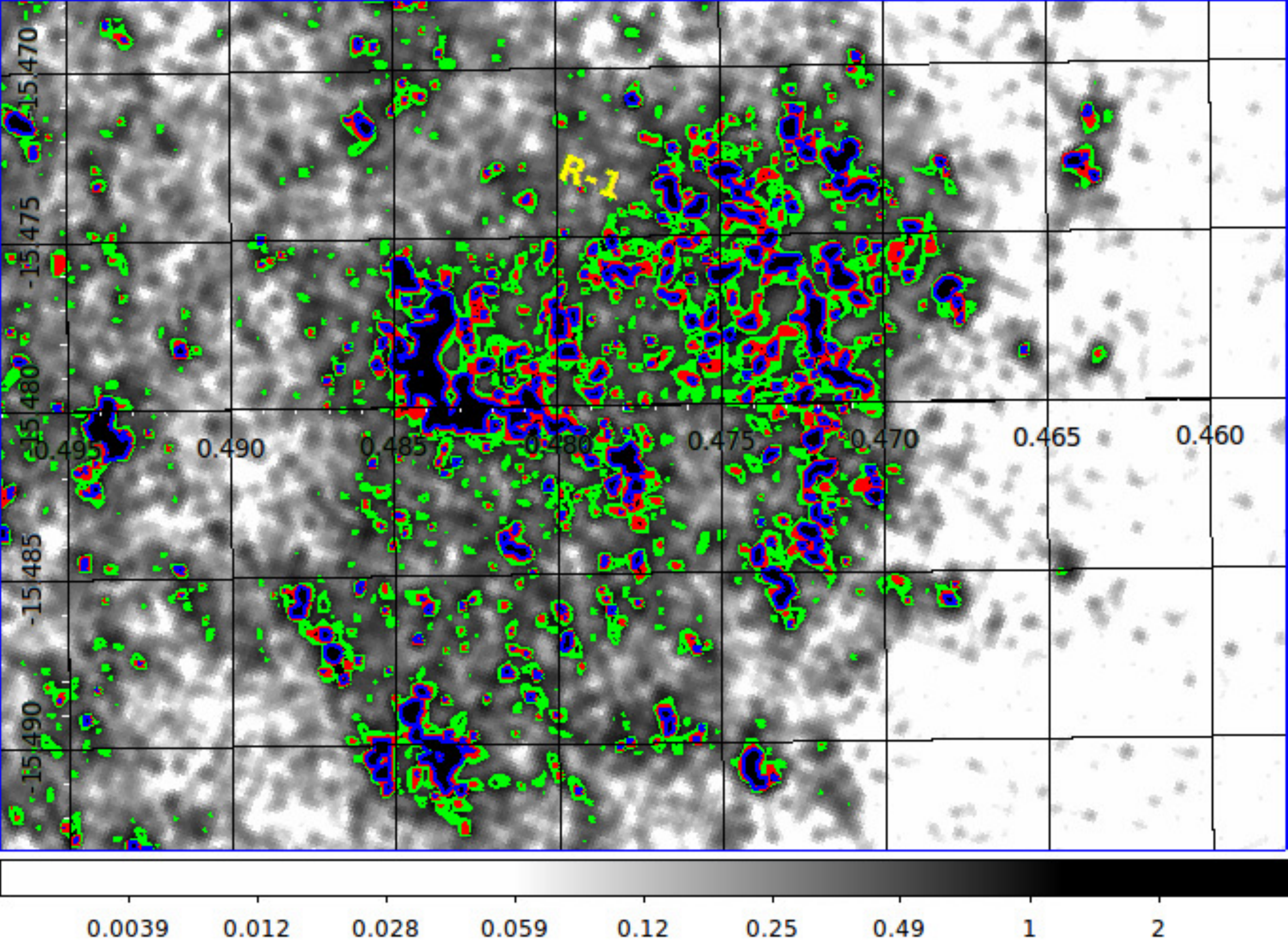}{0.45\textwidth}{(b)}
}
\gridline{\fig{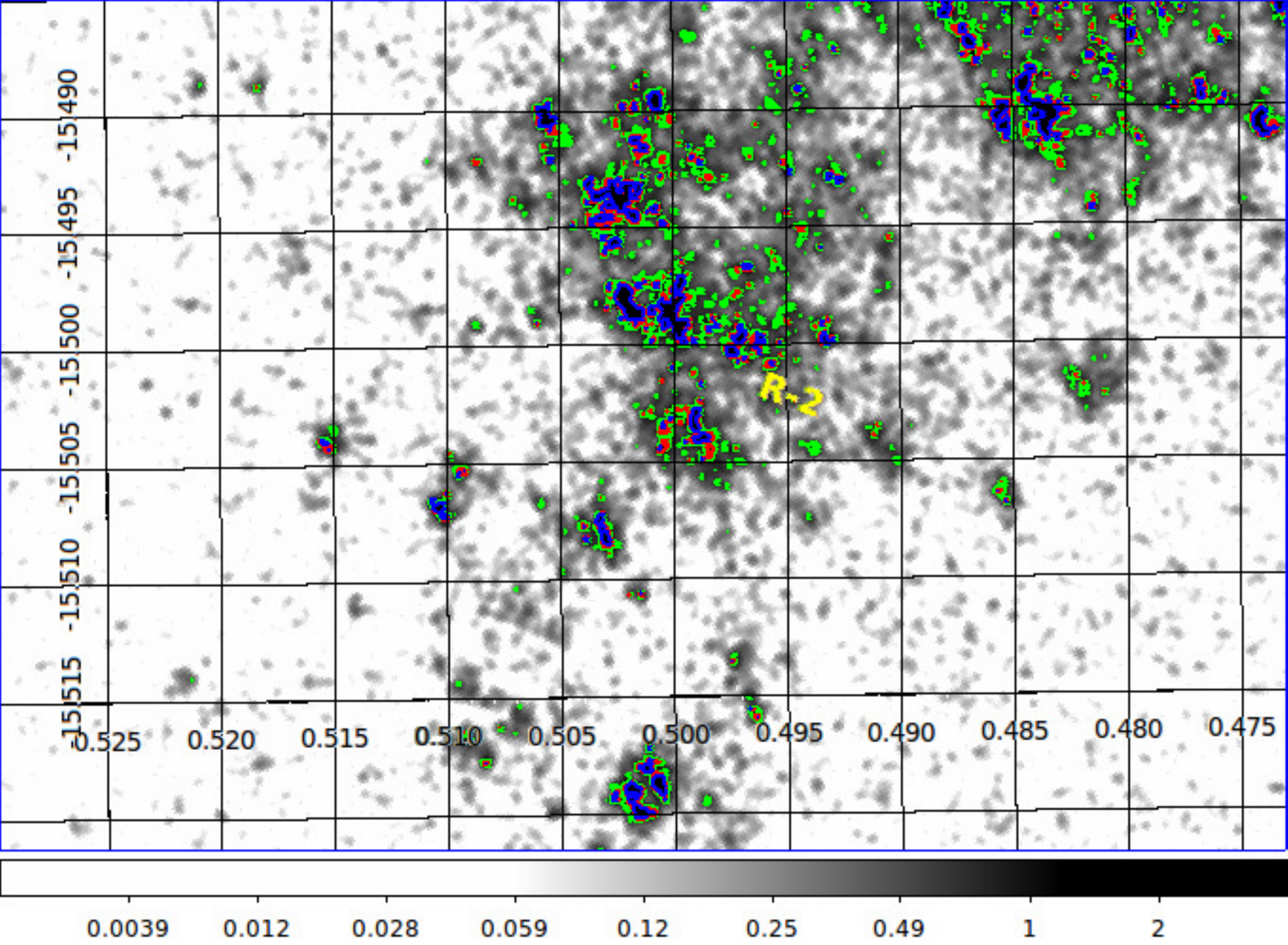}{0.45\textwidth}{(c)}
			\fig{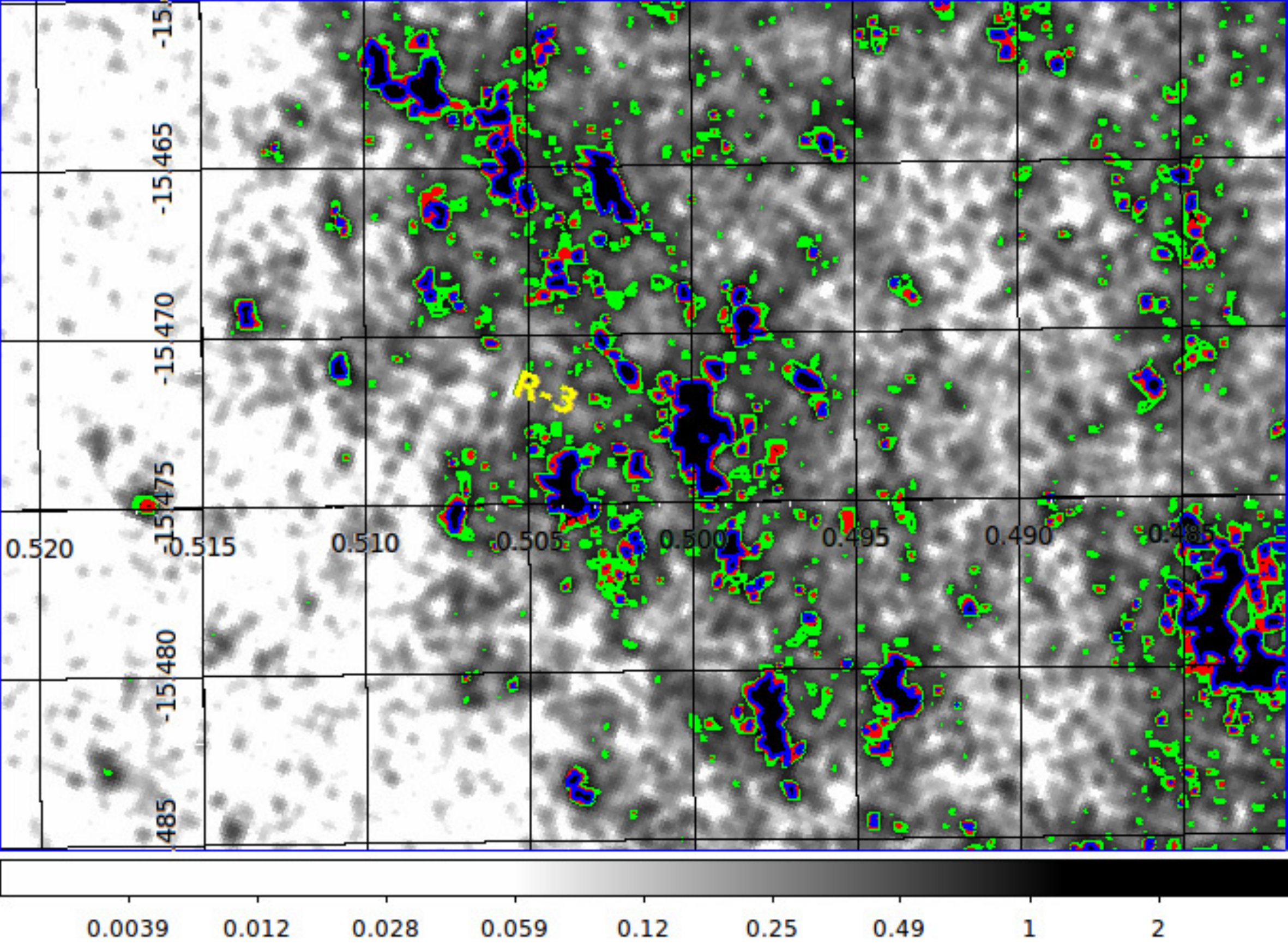}{0.45\textwidth}{(d)}
}
\gridline{\fig{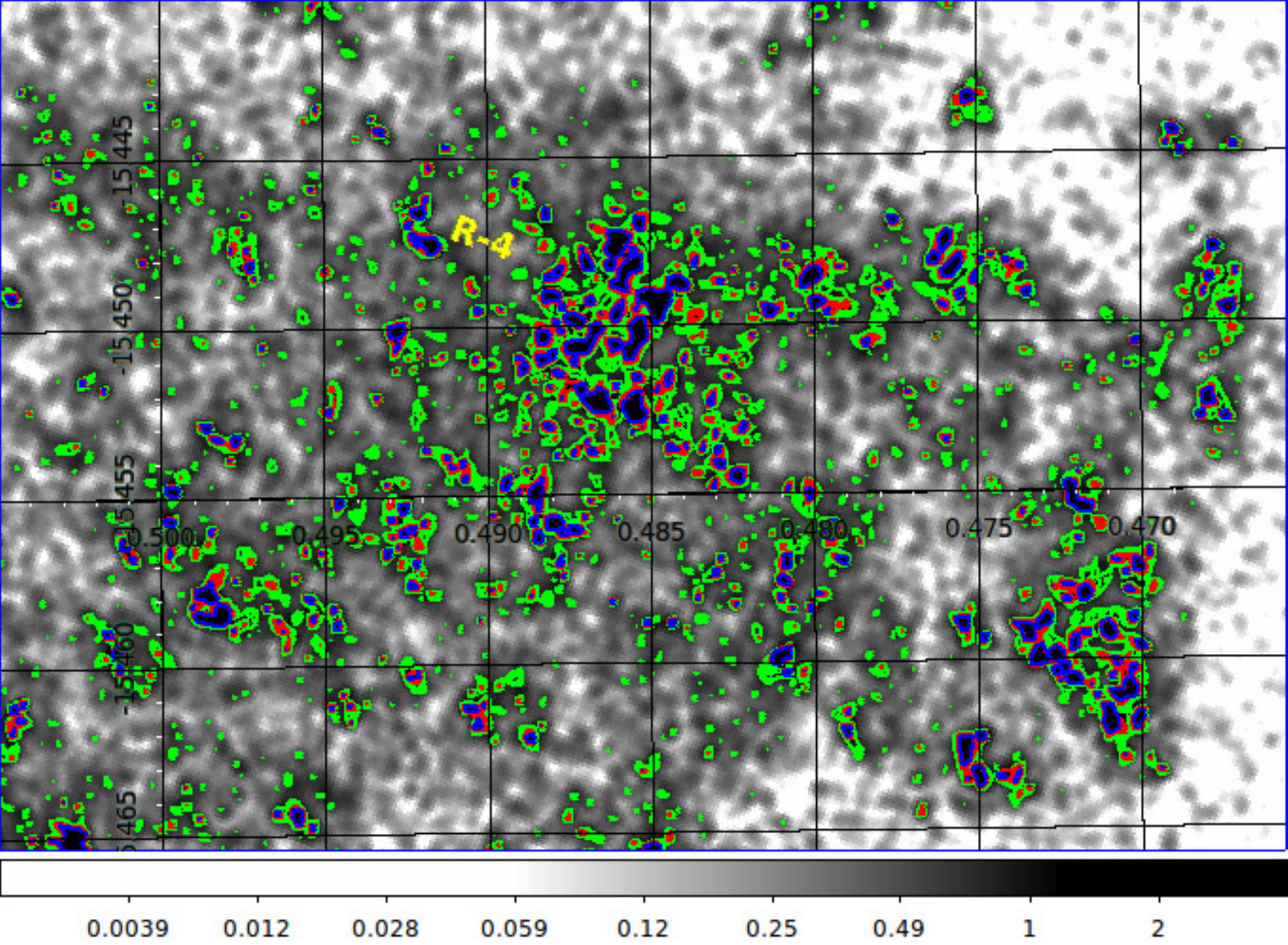}{0.45\textwidth}{(e)}
			\fig{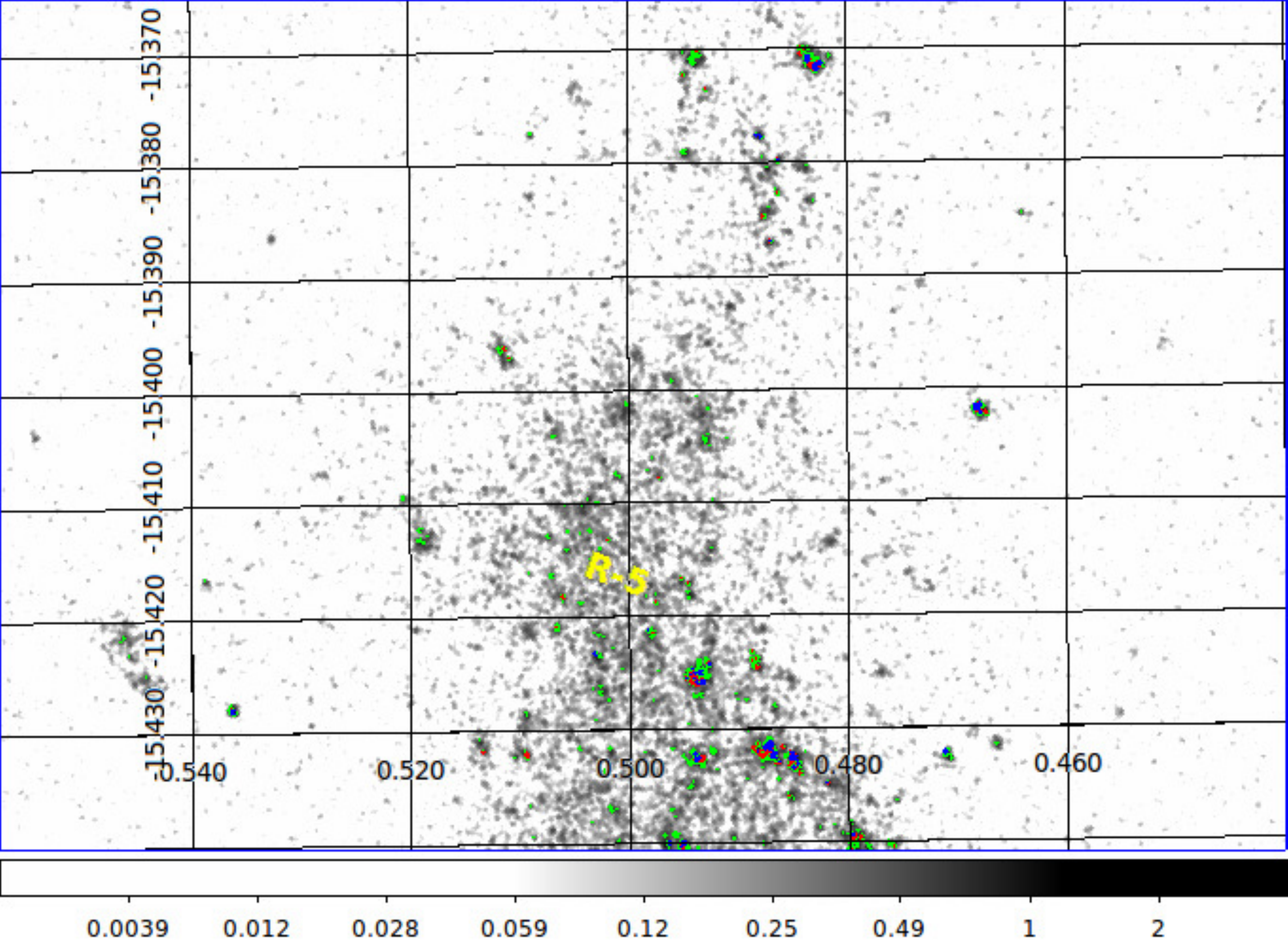}{0.45\textwidth}{(f)}
}
\caption{The (F148W$-$N263M) color map of the galaxy is shown with different plotted contours. Figure (a) shows the whole galaxy along with the specified position of five regions R1, R2, R3, R4 and R5. The blue, red and green contours are generated for different flux limits signifying different temperature ranges as mentioned in Table \ref{F148W/N263M}. The background image shown here is smoothed and the gray scale denotes the logarithmic value of the CPS ratio for the smoothed image. We have shown zoomed in images for regions R1, R2, R3, R4 and R5 in Figure (b), (c), (d), (e) and (f) respectively. The blue contoured regions denote the hotter regions of the galaxy.}
\label{caf2_b4}
\end{figure*}

We first considered (F148W$-$N263M) color map to investigate the temperature profiles of  the star forming regions. Considering the black curve of Figure \ref{kuruz_temp}a, we fixed different upper and lower limits of flux ratio for creating different sets of contour signifying different ranges of temperature which are listed in Table \ref{F148W/N263M}. We adopted a smoothness of 2 pixels ($\sim$ 4 pc), which helps to trace smaller regions as well. We divided the galaxy in five different regions and designated them as R1, R2, R3, R4 and R5. Their locations are shown in Figure \ref{caf2_b4}a. We adopt the same temperature used by \citet{bianchi2012} to classify the regions. The blue contours signify regions with very high temperature (T $>$ 35000 K, corresponding to early O type stars) while red contour indicates an intermediate temperature (25000 K $<$ T $<$ 35000 K, late O type stars) and relatively low temperature (17500 K $<$ T $<$ 25000 K, early B type stars) regions are shown in green contour (Table \ref{F148W/N263M}). Regions which are not covered by any of these three contours are suggestive to have temperatures less than 17500 K. The regions R1 to R5, with overlaid contours are shown in Figure \ref{caf2_b4}b to \ref{caf2_b4}f. Thus the regions covered by these contours are occupied by OB stars. The high temperature regions (blue contour) are found to be enveloped by relatively lower temperature regions (red and green contours) throughout the galaxy. These show the hierarchical structure of actively star forming regions.\\\\

The south-western part (R1) is found to have several high temperature regions (blue contour) of different sizes. Though the regions have irregular shape, an approximate size can be assigned by measuring their longest dimension. The largest high temperature complex identified in R1 is L-shaped, with a north-south extend of
$\sim$ 50 pc, and an east-west extend of $\sim$ 40 pc. The width of the region is in the range, $\sim$20-30 pc. This is the largest massive complex in WLM. We also found nearly 10 regions of size $>$ 20 pc. The smaller regions ($<$ 10 pc) are more scattered and large in number, with smallest of them with sizes $\sim$ 4 pc. The red contours signifying intermediate temperature are found to connect several hot regions (blue contour) and form knotted structures. It is also noticed that the number of smaller ($\sim$4 pc) low temperature regions are more than similar size hotter regions. The nature of temperature gradient around each hot star forming region (blue contour) is clearly depicted by the red and green contours present around them. \\\\
The region R2 is assigned to the southern part of the galaxy. We identify three high temperature regions of size $\sim$ 30 pc, and a large number of smaller regions with sizes $<$ 10 pc. One clumpy hot region is identified at the extreme south of the galaxy, which appears to be almost detached from the main body of the galaxy. The eastern part of the galaxy (R3) is found to have more number of larger-sized high temperature regions, with  8 complexes of size larger than 20 pc and one region extending up to $\sim$ 50 pc in length. These 8 complexes are spatially isolated from each other and distributed along the south-west to north-east direction. Around some of these large complexes, we noticed smaller ($<$ 10 pc) hot regions which are connected by red contours. The scenario of R4 (north and north-western part) region is a little different. Though the region has a larger spatial extent than the rest, we do not detect any large structure with very high temperature, rather an abundance of structures scattered all around. We also identify two locations (one along north and another along west) with increased number of scattered structures. The extreme northern part of the galaxy (R5) does not show any high temperature region except two or three isolated hot star forming clumps.\\\\
Therefore the star forming regions of WLM are found to have an overall hierarchical structure with hottest cores surrounded by relatively less hotter regions. It is also clear that the young star forming regions are fragmented and clumpy in nature. 

\subsection{(F148W$-$N245M) color map}
\begin{figure*}
\centering
\gridline{\fig{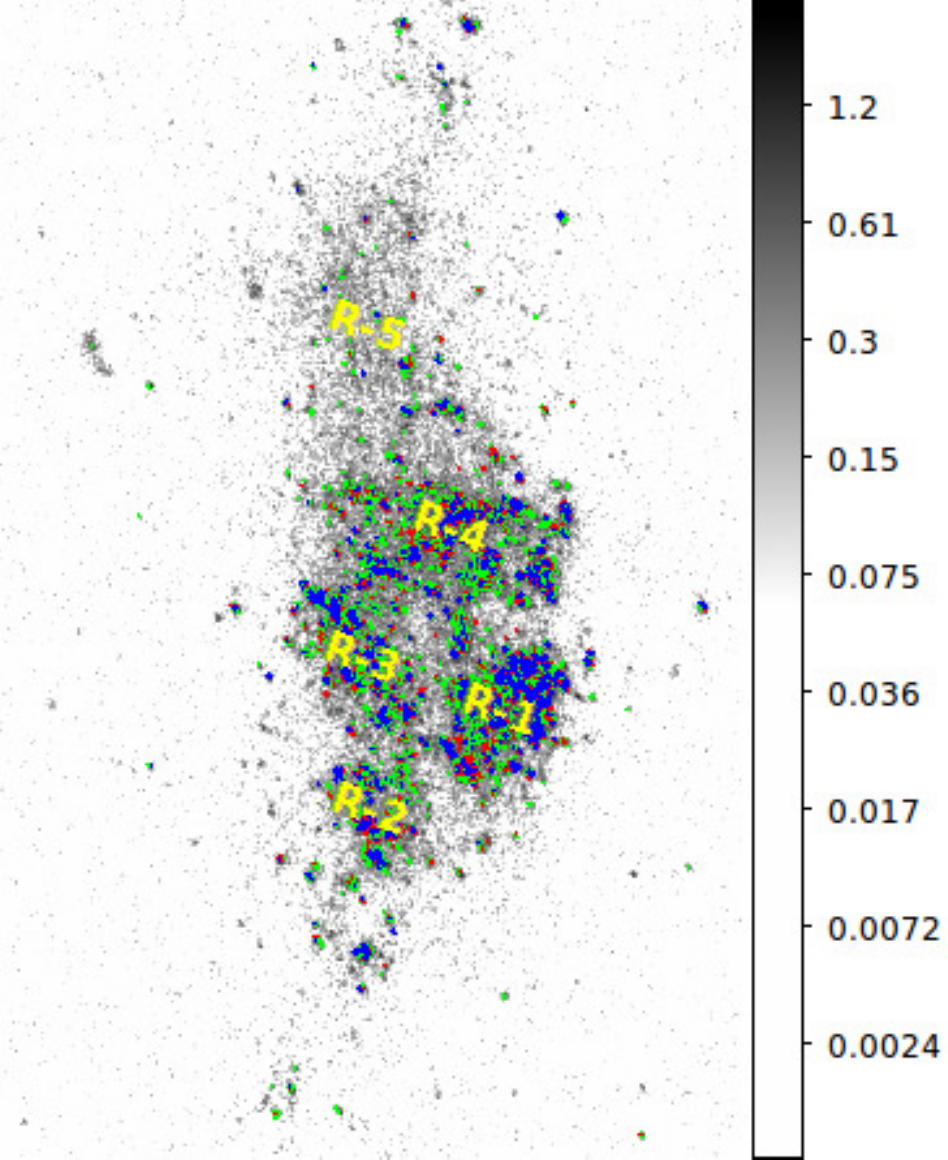}{0.35\textwidth}{(a)}
			\fig{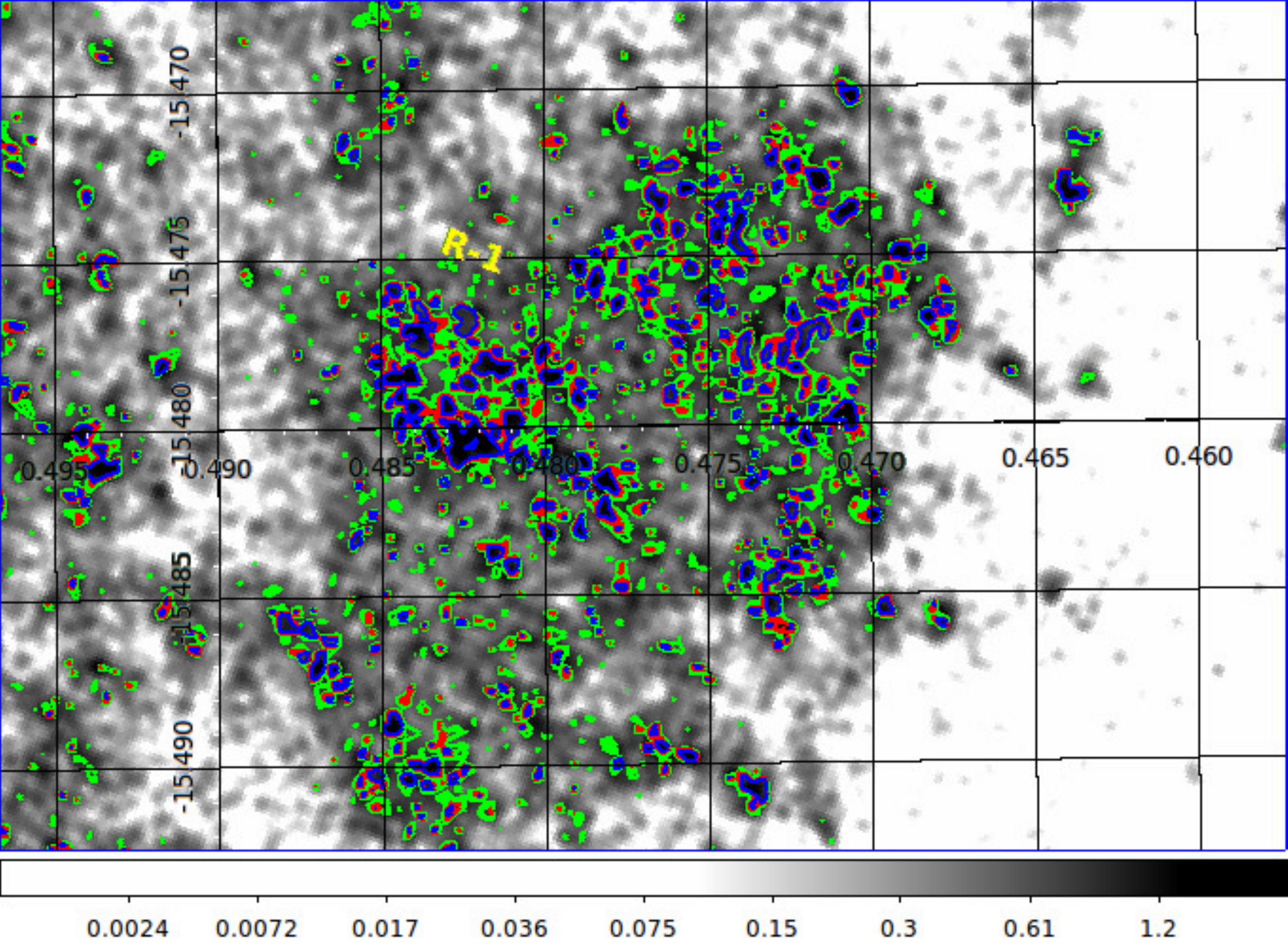}{0.45\textwidth}{(b)}
}
\gridline{\fig{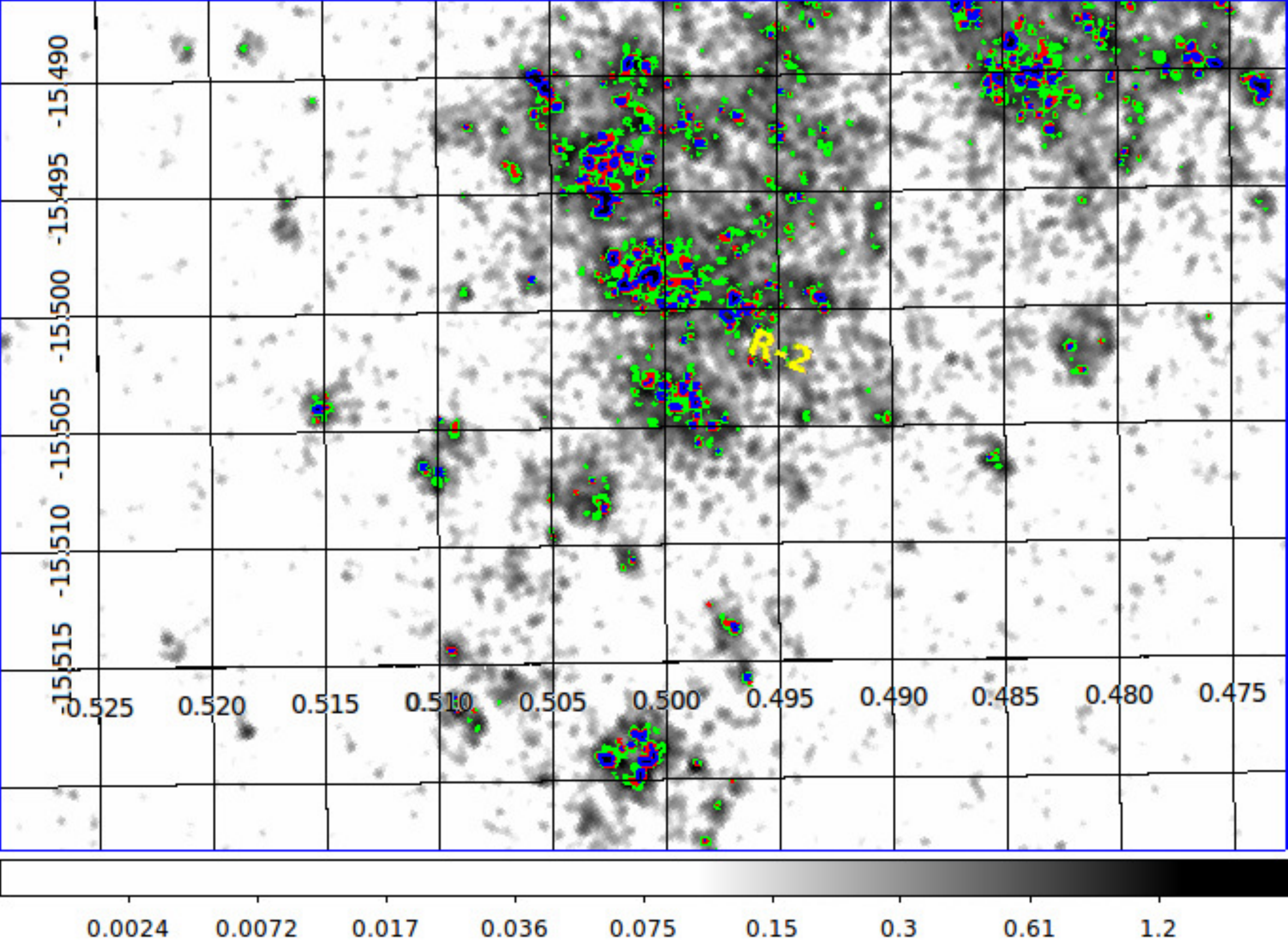}{0.45\textwidth}{(c)}
			\fig{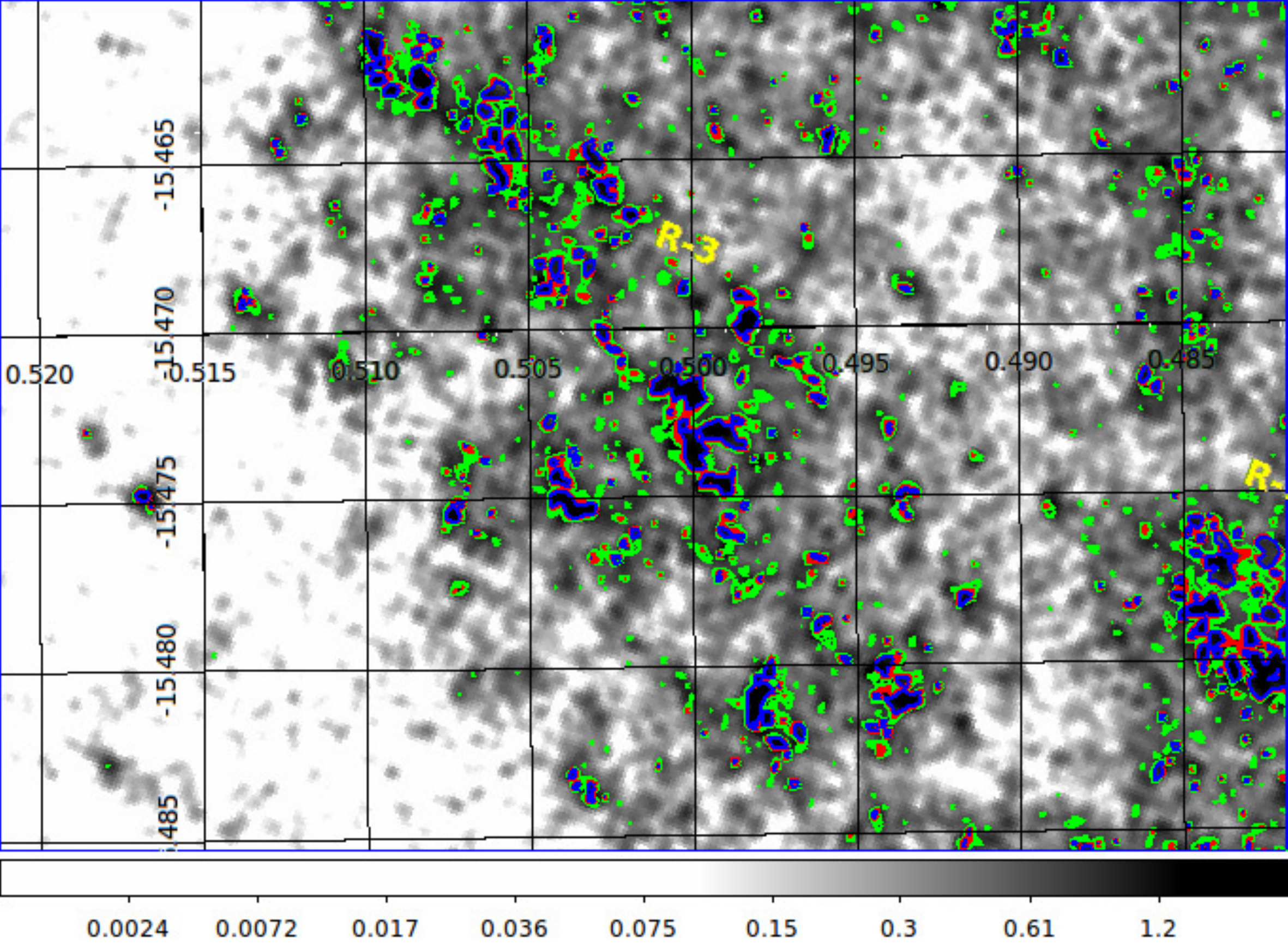}{0.45\textwidth}{(d)}
}
\gridline{\fig{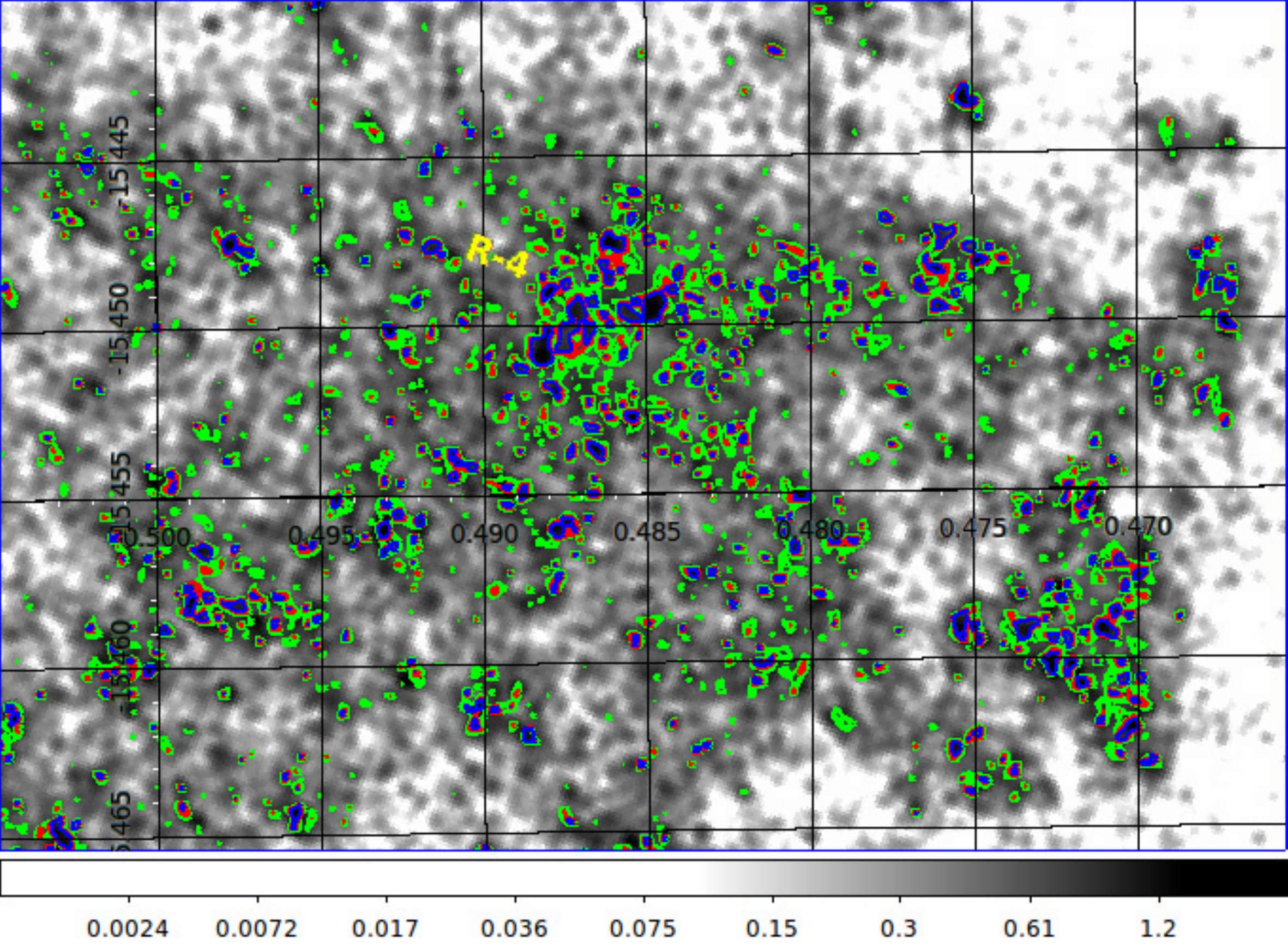}{0.45\textwidth}{(e)}
			\fig{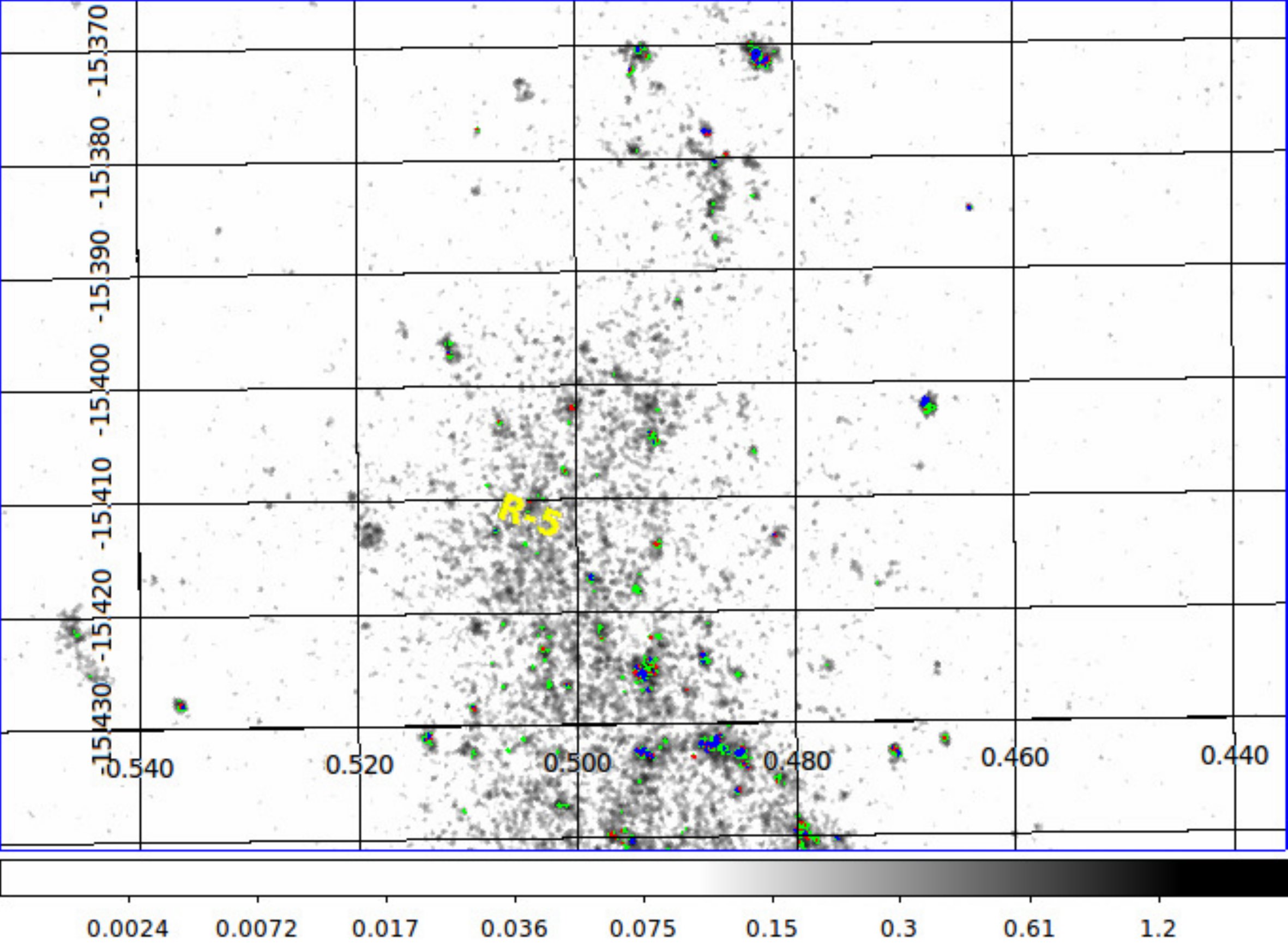}{0.45\textwidth}{(f)}
}
\caption{The (F148W$-$N245M) color map of the galaxy is shown with different plotted contours. Figure (a) shows the whole galaxy along with the specified position of five regions R1, R2, R3, R4 and R5. The blue, red and green contours are generated for different flux limits signifying different temperature ranges as mentioned in Table \ref{F148W/N245M}. The background image shown here is smoothed and the gray scale denotes the logarithmic value of the CPS ratio for the smoothed image. We have shown zoomed in images for regions R1, R2, R3, R4 and R5 in Figure (b), (c), (d), (e) and (f) respectively. The blue contoured regions are found to be more fragmented than those in Figure \ref{caf2_b4}.}
\label{caf2_b13}
\end{figure*}

In order to explore the effect of differential extinction inside the galaxy and to study the detected hot star forming regions further, we performed the same exercises for color map (F148W$-$N245M) (i.e. F148W/N245M image). The model generated diagram for tracing the temperature in (F148W$-$N245M) color map is shown in Figure \ref{kuruz_temp}b (blue curve). 
The UV color image with the overlaid contours are shown in Figure \ref{caf2_b13}a. The expanded view of R1, R2, R3, R4 and R5 regions are shown in Figure \ref{caf2_b13}b to \ref{caf2_b13}f. It is natural to expect the shape and distribution of the identified hot star forming regions to be similar in nature in both the color maps. We notice that, though the overall distribution of hot star forming regions in both the color maps matches in general, a closer look reveals a more fragmented structure in the (F148W$-$N245M) color map. The larger complexes detected in (F148W$-$N263M) color maps are noticed to split into smaller regions in (F148W$-$N245M) color map.
This could happen because of two reasons. The differential behaviour of 2175 $\AA$ bump inside the galaxy can change the effective extinction value which affects the observed flux value in N245M filter. This in turn can produce the structural differences of star forming regions between the two color maps. The same effect can also be produced by a clumpy distribution of hot stars, which have more flux in N245M filter. Thus, we conclude that either the young and hot stellar groups or the dust or both of them present in WLM have a clumpy distribution. 

\subsection{Correlating with H$\alpha$ and H$~$I maps}
\begin{figure}
\begin{center}
\includegraphics[width=3.3in]{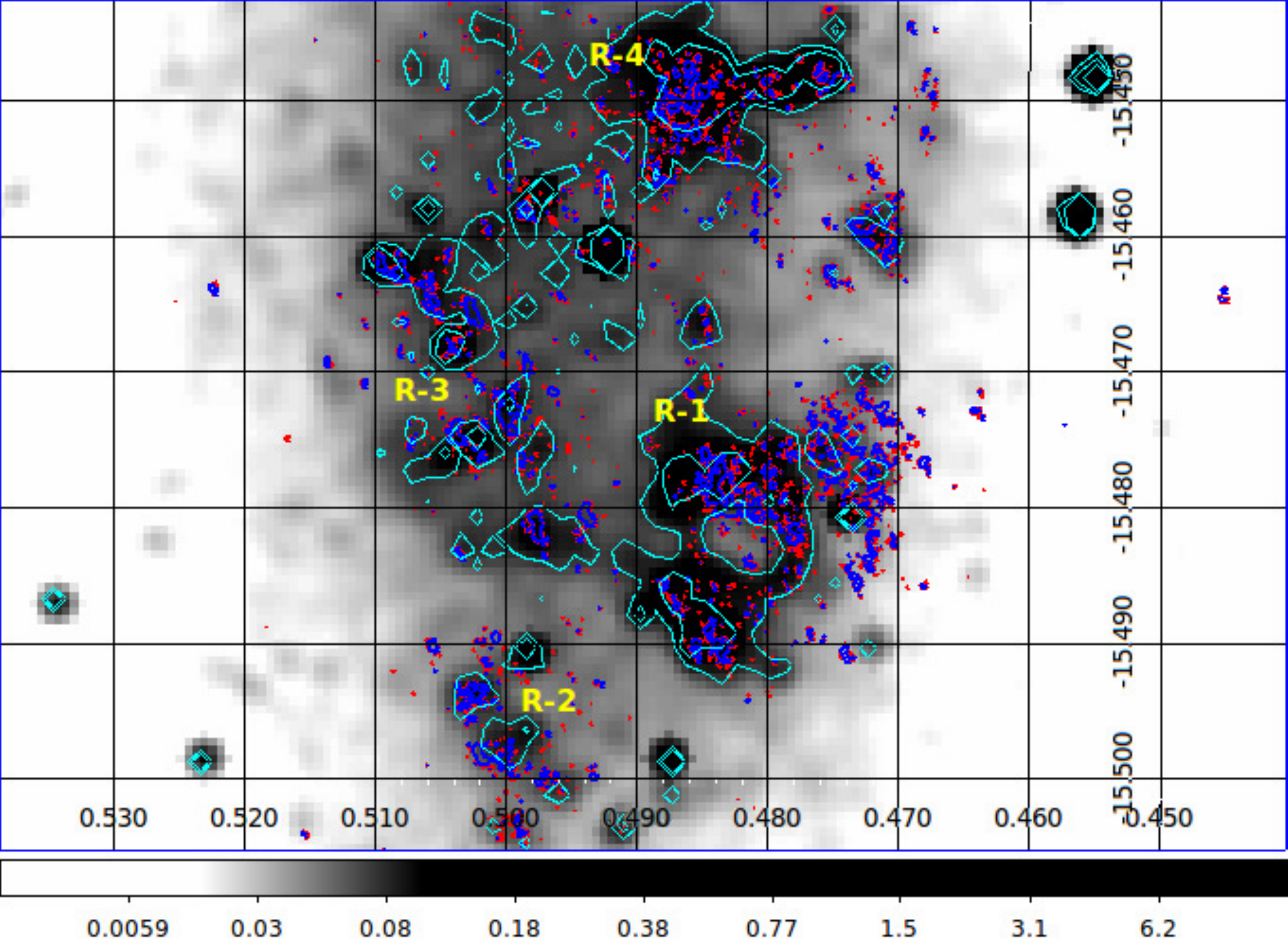} 
 \caption{H$\alpha$ image of the galaxy WLM, covering regions R1, R2, R3 and R4, is shown in the background where the gray scales signifies the logarithmic value of CPS. The cyan contours signify the H$\alpha$ emitting regions of the galaxy. The plotted blue and red contours are same as shown in Figure \ref{caf2_b4}a. The UV detected hot regions show a good spatial correlation with the H$\alpha$ emitting regions of the galaxy. }
 \label{Halpha}
 \end{center}
 \end{figure}

We correlate the spatial distribution of UV detected high temperature regions in the color map (i.e. (F148W$-$N263M)) with H$\alpha$ and H$~$I image of the galaxy. Since H$\alpha$ radiation mainly comes from gas ionised by massive and hot OB type stars, it is expected that the UV detected high temperature regions will show a good spatial correlation with H$\alpha$ emitting regions. In Figure \ref{Halpha}, along with the contours (blue and red) created in (F148W$-$N263M) color map, we have shown the H$\alpha$ emitting regions of the galaxy in cyan contours. Almost all of the high temperature regions detected in R1, R2, R3 and R4 appear inside the cyan contours, which indeed signifies a good spatial correlation between them. In the R1 region, we notice a hole in the H$\alpha$ map which does not have any detected UV hot region near its center, though we detect a few smaller hot regions within the hole. We identify a large number of hot regions located to the north of the hole, with reduced number of such regions in the southern side. We identify several clumpy and knotted hot regions towards west of R1 region with very less H$\alpha$ emission. In R2, there are two major hot clumps which also show H$\alpha$.
All the hot regions detected in R3 region correlate well with H$\alpha$ emitting regions. The shapes of the H$\alpha$ contours also match with the overall distribution pattern of high temperature regions. In R4 region, the detected clumps of hot regions are enveloped by H$\alpha$ emission. The elongation of the central H$\alpha$ contour in R4 along west and south is also traced well by the hot regions. The overall good spatial correlation between H$\alpha$ emitting regions and UV detected hot complexes confirms the presence of massive O and early B type stars in these complexes.\\
 
\begin{figure}
\begin{center}
\includegraphics[width=2.8in]{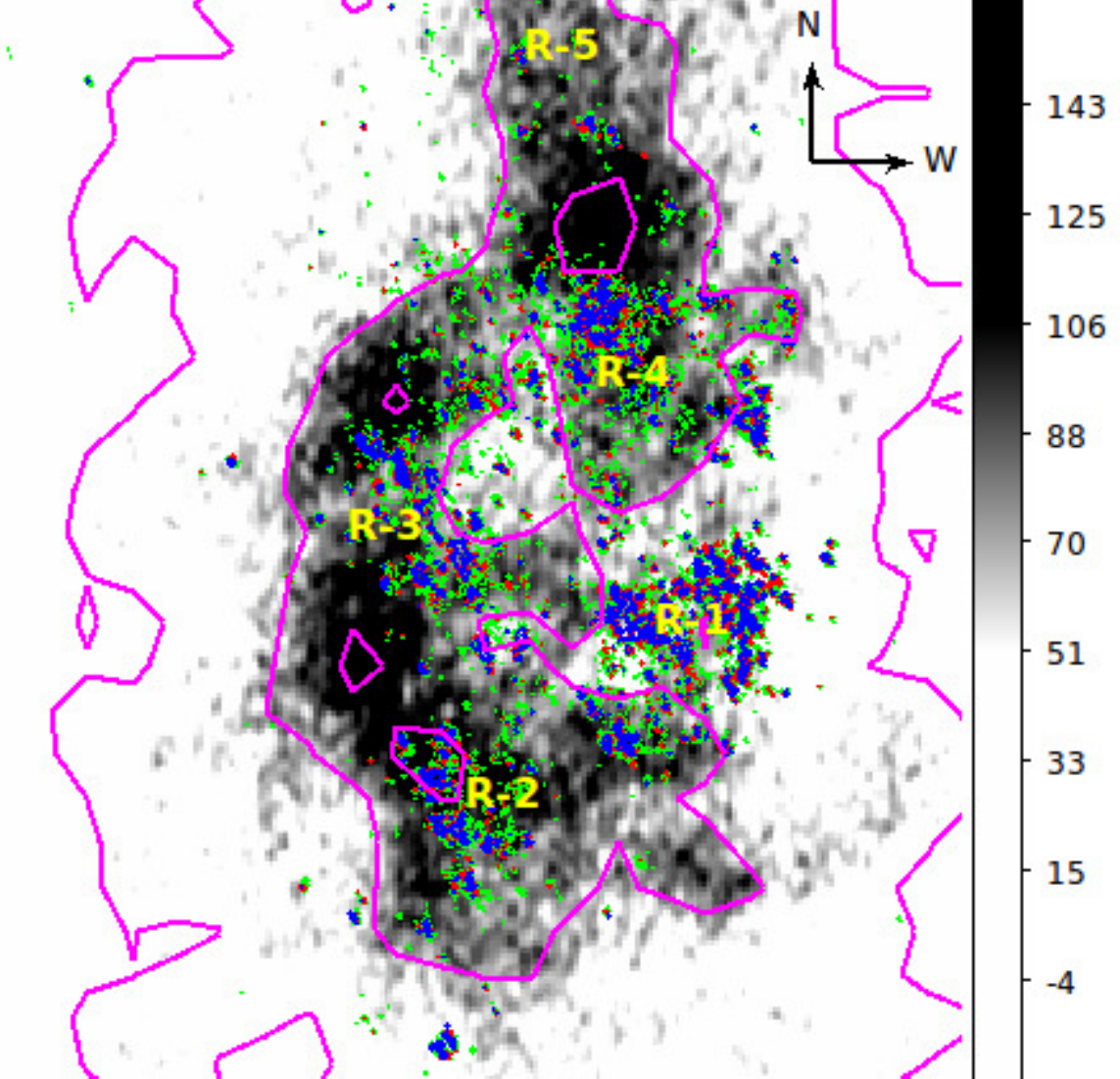} 
 \caption{The background image shows the H$~$I density distribution of the galaxy WLM. The hook like magenta contour is the over density pattern reported by \citet{kepley2007}. The plotted blue, red and green contours are same as shown in Figure \ref{caf2_b4}a. R1 has several UV detected hot star forming regions with less H$~$I density. Other regions (R2, R3, R4 and R5) are found to have dense H$~$I gas.}
 \label{HI}
 \end{center}
 \end{figure}

 Further, we spatially correlate these hot star forming regions with neutral hydrogen (H$~$I) density map of the galaxy. Since stars are mainly formed from molecular clouds,
the presence of hydrogen gas would support the possibility of recent star formation. \citet{kepley2007} identified a hook like over density pattern in the H$~$I distribution as shown in Figure \ref{HI}. The hook shaped magenta contour is found to cover almost all the high temperature regions except those in R1. Therefore, R1 region is found to have many UV detected hot star forming clumps showing H$\alpha$ emission, (except some part in west - Figure \ref{Halpha}) but with a relatively low H$~$I density. This may be the result of a vigorous recent star formation in R1, where the H$~$I gas is ionised/driven away after the star formation. Other hot regions in R2, R3 and R4 are found to have a high column density of H$~$I along with H$\alpha$ emission.

\subsection{Correlating with HST detected hot stars}
\begin{figure*}
\centering
\gridline{\fig{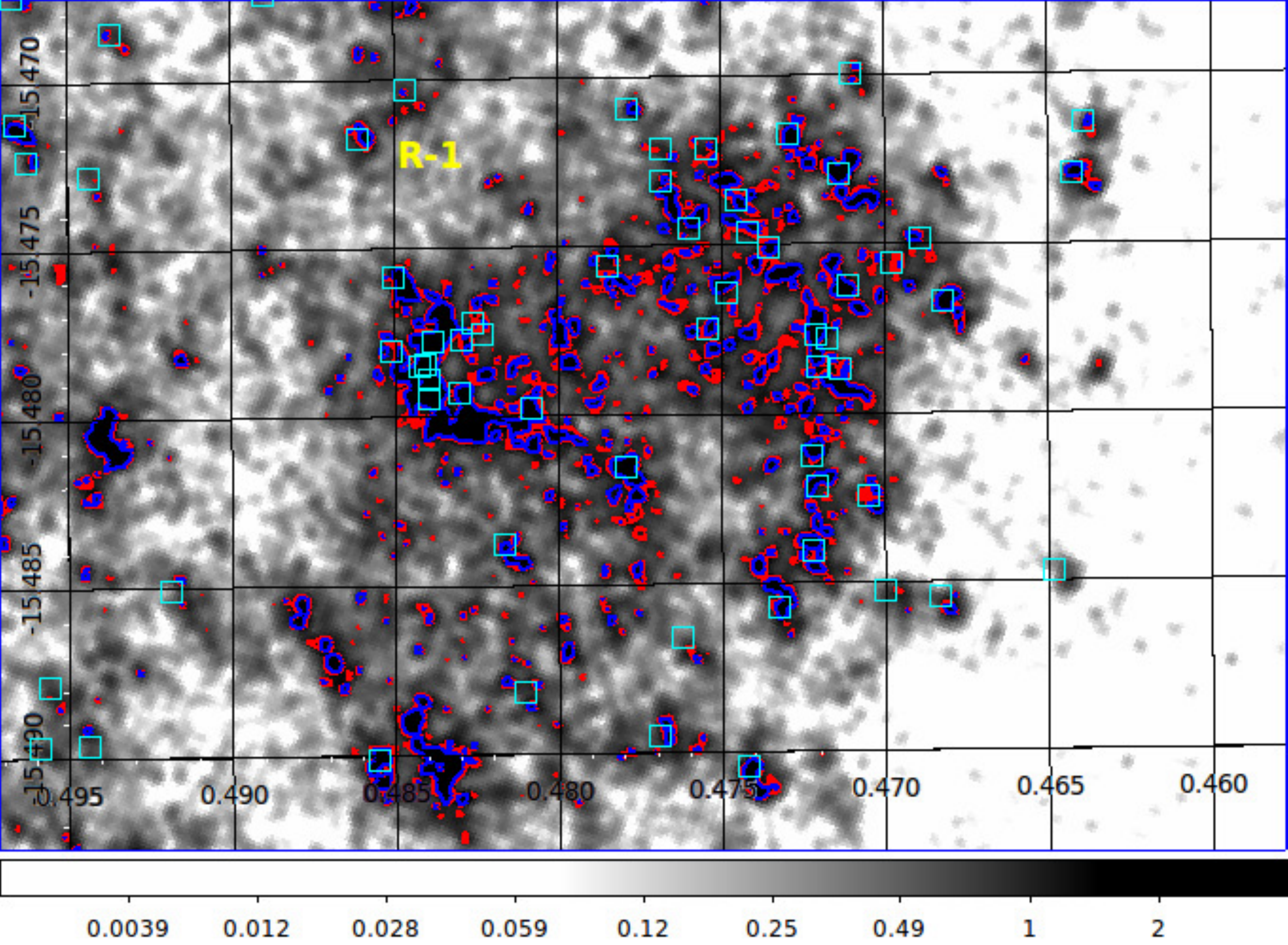}{0.47\textwidth}{(a)}
			\fig{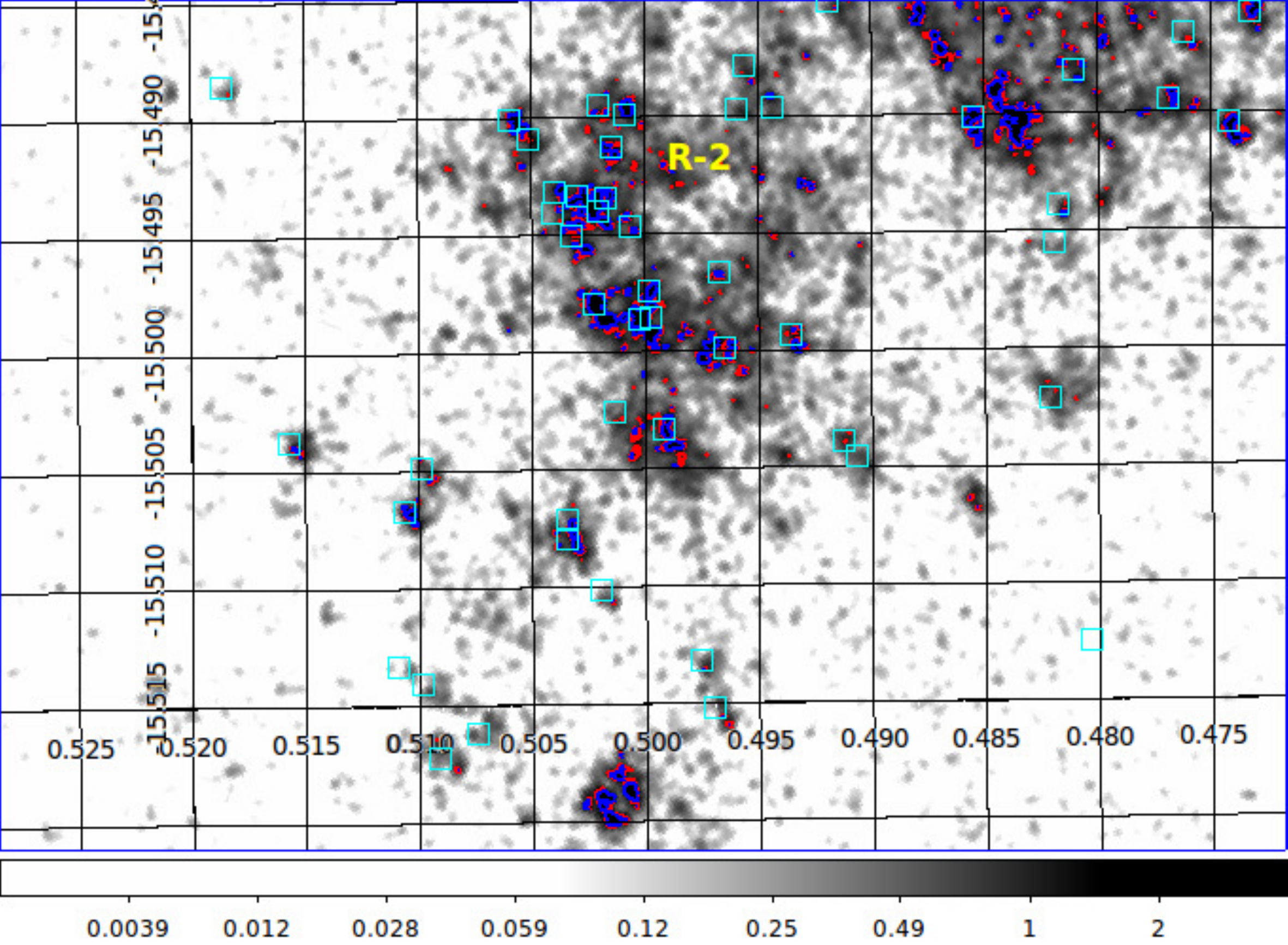}{0.47\textwidth}{(b)}
}
\gridline{\fig{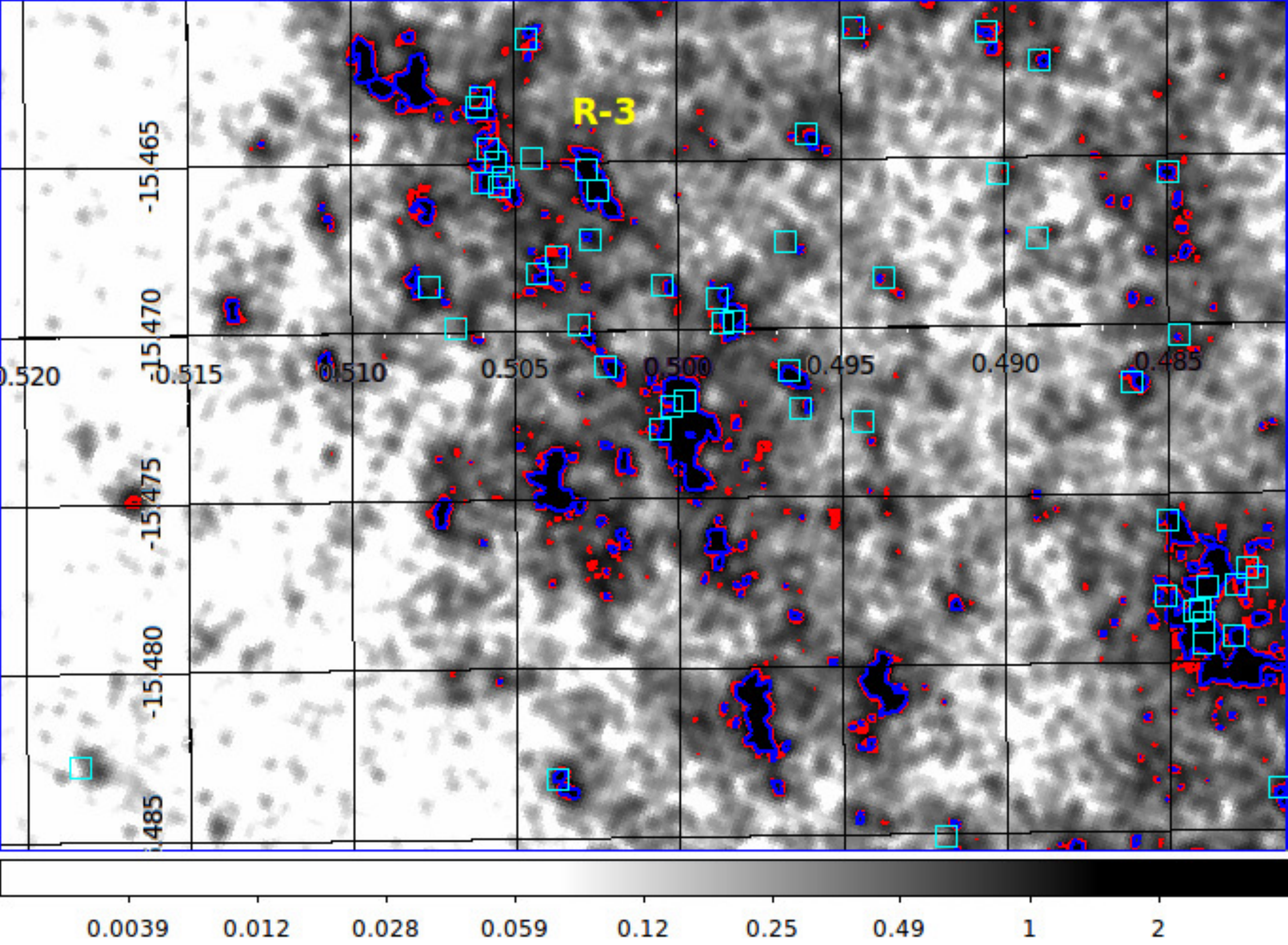}{0.47\textwidth}{(c)}
			\fig{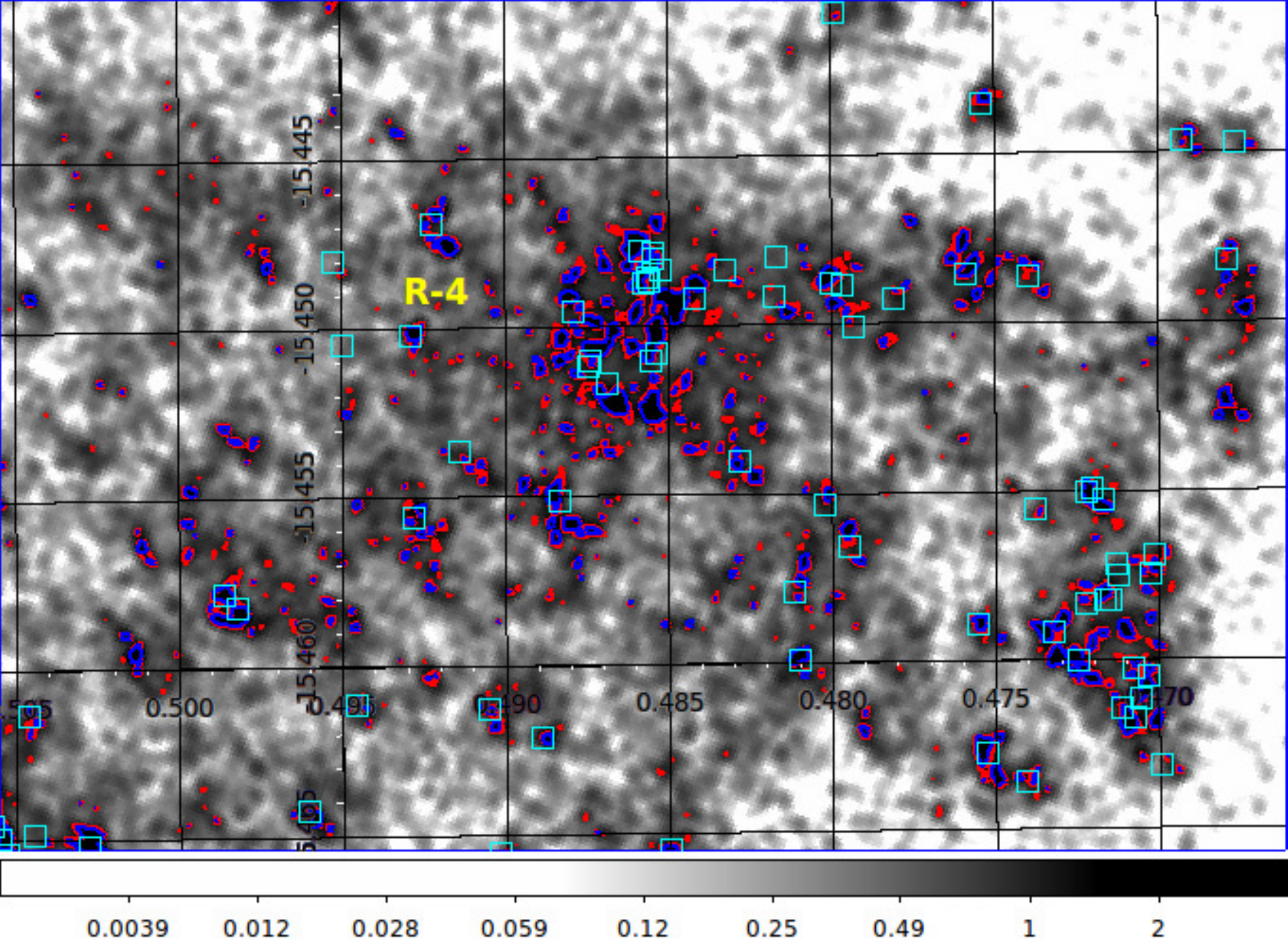}{0.47\textwidth}{(d)}
}
\caption{F148W/N263M smoothed image is shown with same blue and red contours as plotted in Figure \ref{caf2_b4}. The HST detected massive stars (cyan box) are overlaid on the regions R1, R2, R3 and R4 as shown in figures (a), (b), (c) and (d) respectively.}
\label{caf2_b4_hst}
\end{figure*}

 \citet{bianchi2012} observed the galaxy WLM with six HST broadband filters (F170W, F255W, F336W, F439W, F555W, F814W) and noticed a richness of stars with age younger than 10 Myr. The bluest band used in their observation is F170W with a bandwidth of 1200 - 2400 $\AA{}$. Stars which are bright in F170W band are expected to be hot, young and relatively massive. We spatially correlated the hot regions detected in our study with stars that are brighter than 19 magnitude in F170W band from their photometric catalog. Excluding some part of R3, the regions R1, R2, R3 and R4 are almost fully covered by the three HST fields observed in their study. In Figure \ref{caf2_b4_hst} we have overlaid these stars on top of (F148W$-$N263M) color map. The detected stars (cyan box) show a good spatial correlation with the hot star forming regions (blue and red contour) of the galaxy. This confirms that massive young stars are co-located with the high temperature star forming regions identified in our study.
 
 \subsection{Mass estimation of compact star forming regions}
\begin{figure}
\begin{center}
\includegraphics[width=3.7in]{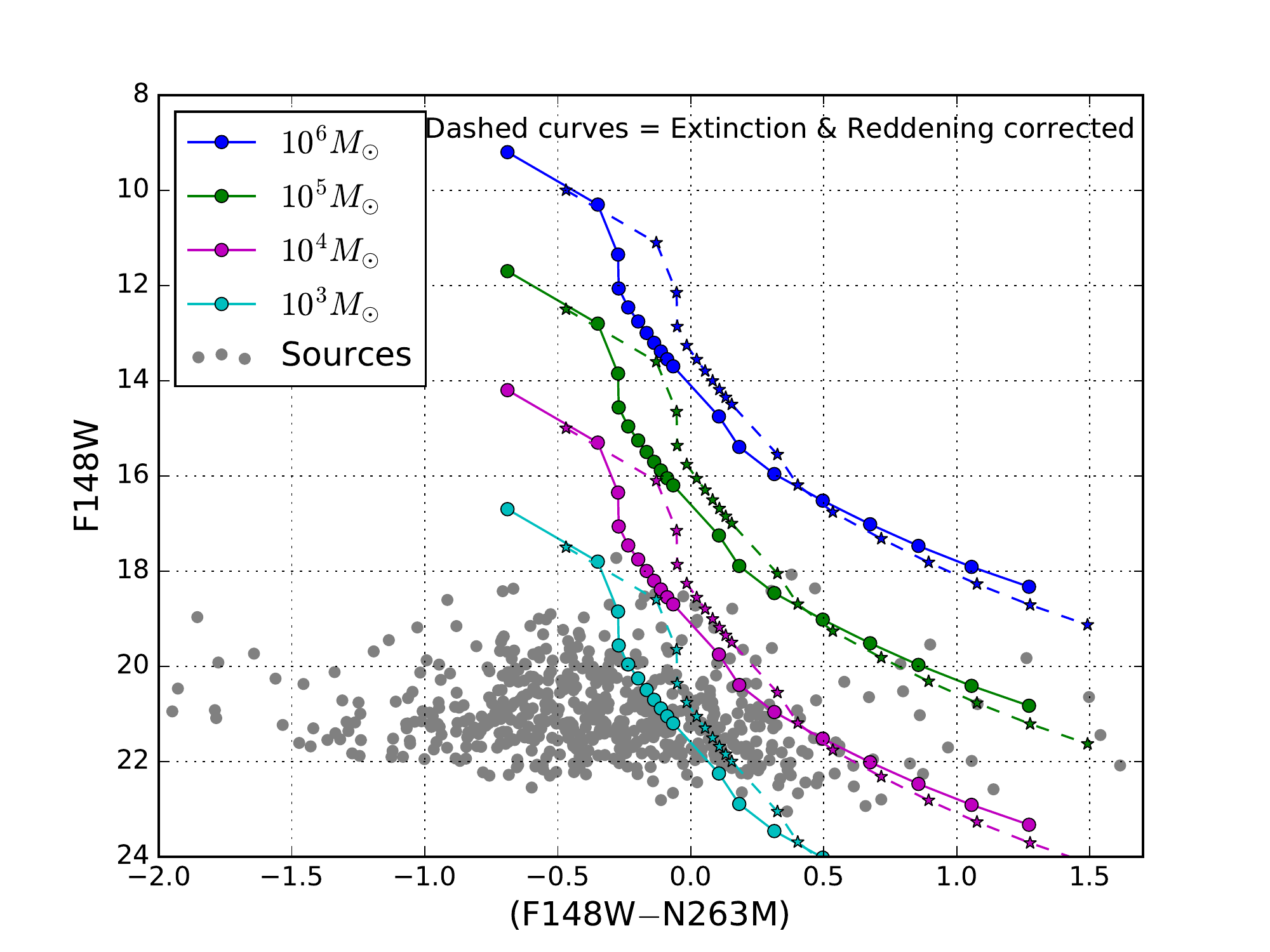} 
 \caption{The identified compact star forming regions are over plotted (gray filled circles) on starburst99 model curves.}
 \label{cmd_caf2_b4}
 \end{center}
 \end{figure}

In the UVIT images, some of the star forming regions were found to appear like point sources, due to their small sizes. As these are a good number of them, many of these are likely to be part of WLM, though a small fraction could belong to the background. We assumed all the detected point sources to be the compact star forming regions of the galaxy and performed PSF photometry for estimating their magnitudes. In order to estimate mass, we considered the filters F148W and N263M for which we have simulated the model diagram in Figure \ref{cmd_caf2}. After cross matching the individually detected star forming regions in F148W and N263M filters, we identified 590 common regions.  
In Figure \ref{cmd_caf2_b4}, we plotted the observed star forming regions (grey points) along with the model curves as previously shown in Figure \ref{cmd_caf2} to estimate the mass of these regions. Most of the detected star forming regions have colors within the model range, whereas we do detect a significant number of sources with bluer colors. These may be due to background sources and hence we consider only those which appear within the model limit. The majority of the compact star forming regions are found to have mass M $< 10^3 M_{\odot}$ with some of them between $10^3 M_{\odot}$ and $10^4 M_{\odot}$. A few regions are found to have M $> 10^5 M_{\odot}$, but could be older as indicated by their color. Therefore the galaxy WLM is found to have a large number of young low mass compact star forming regions with M $< 10^3 M_{\odot}$.

 \subsection{Correlating with CO observations}
\begin{figure}
\begin{center}
\includegraphics[width=3.2in]{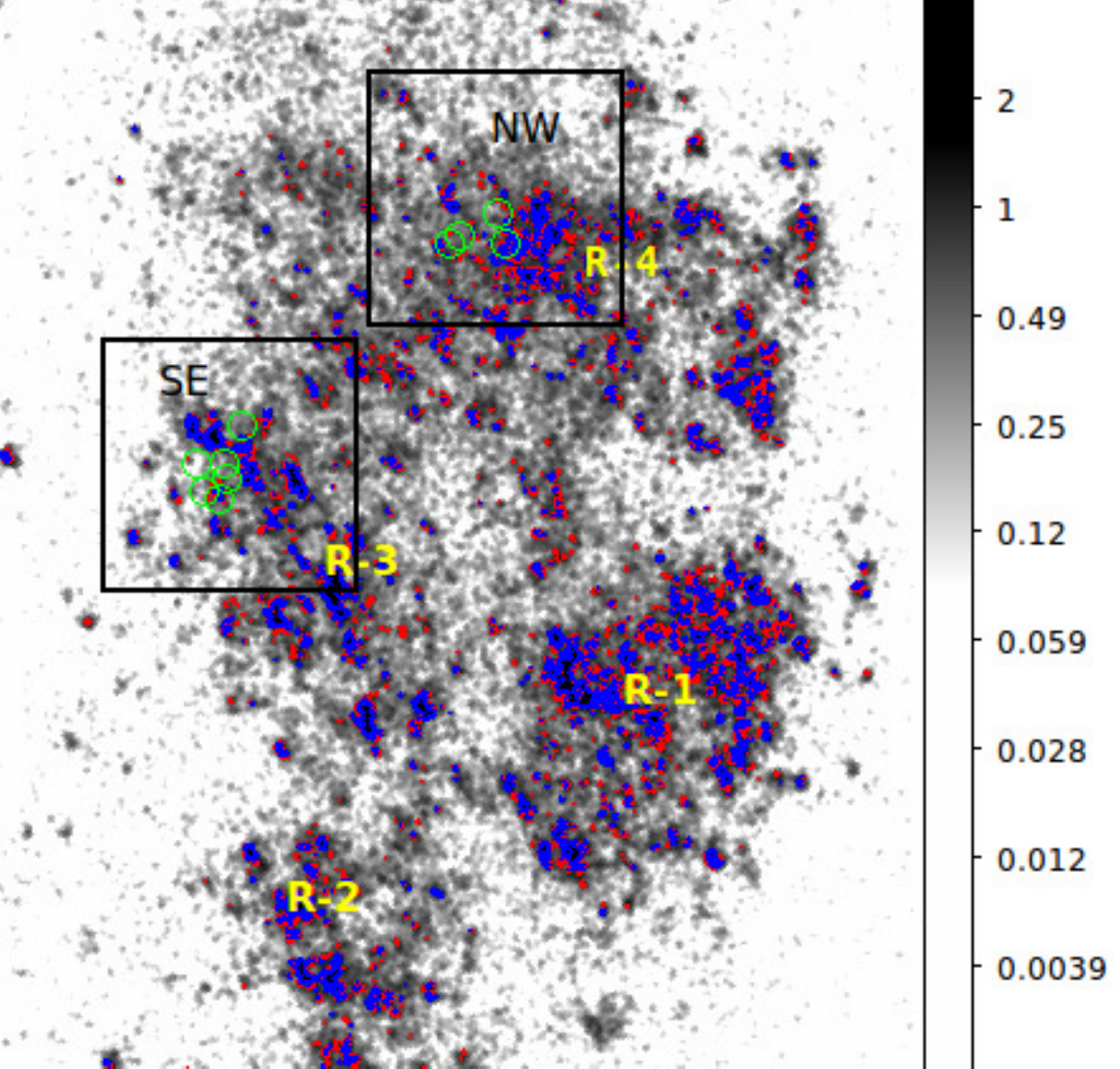} 
 \caption{The smoothed F148W/N263M image is shown in the background with the logarithmic gray scale. The blue and red contours signify the hot regions of the galaxy as in Figure \ref{F148W/N263M}a. Two black squares (one in north-west (NW) and another in south-east (SE)) are the two 1 arcmin$^2$ ALMA fields observed by \citet{rubio2015} for detecting CO clouds. The green circles shown inside the boxes are the position of detected CO clouds.} 
 \label{co_full}
 \end{center}
 \end{figure}

\begin{figure}
\begin{center}
\gridline{\fig{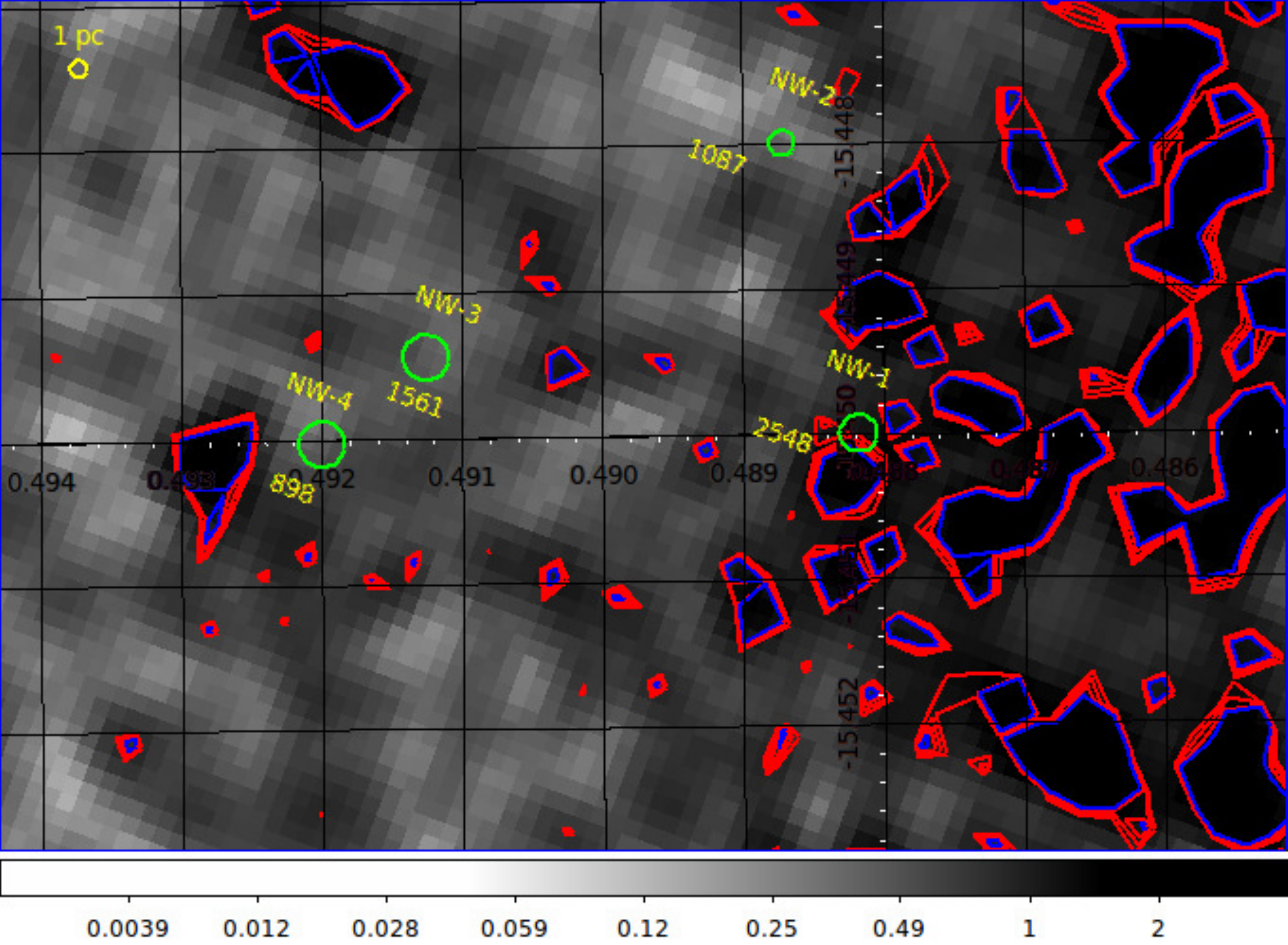}{0.45\textwidth}{(a)}
}
\gridline{\fig{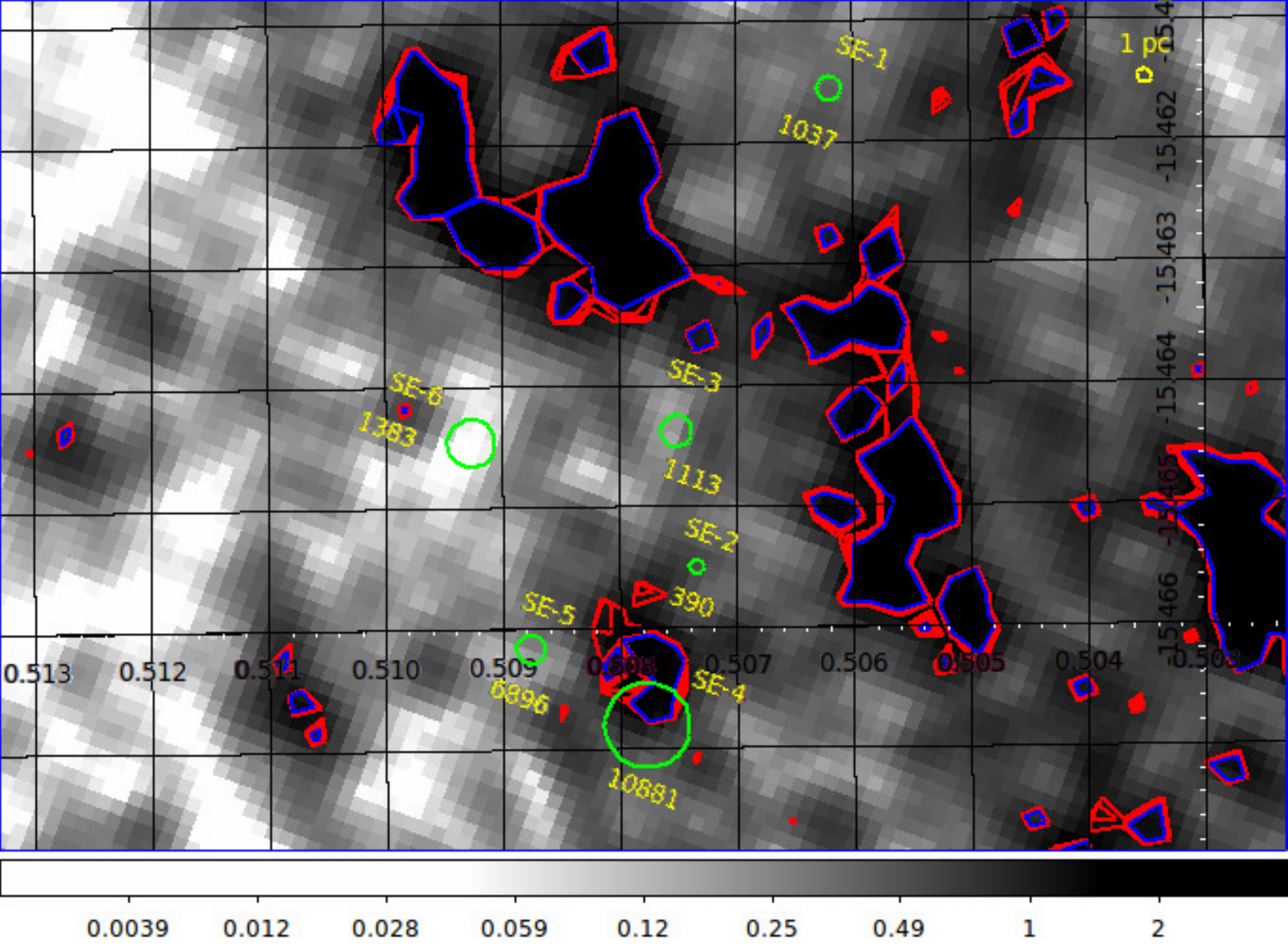}{0.45\textwidth}{(b)}
}
 \caption{Figure (a) and (b) show the zoomed in view of the regions (inside two black boxes of Figure \ref{co_full}) where CO detection is found in north-west (NW) and south-east (SE) field respectively. The background image and contours are same as mentioned in Figure \ref{co_full}. The CO clouds are shown in green circles where the size of the circles signifies the size of the cloud with respect to the 1 pc reference circle shown in each figure. The clouds are labelled as per \citet{rubio2015}. The numbers shown near each green circle signify its virial mass in $M_{\odot}$.}
 \label{co_zoom}
 \end{center}
 \end{figure}

 Since molecular clouds are the sites of star formation, we correlate our results with a recent CO observations of WLM. With the help of ALMA, \citet{rubio2015} observed two 1 arcmin$^2$ field in WLM with a spatial resolution of 6.2 pc $\times$ 4.3 pc which is comparable with UVIT. They identified 10 CO clouds with an average radius of 2 pc and an average virial mass of 2$\times$10$^3$ $M_{\odot}$. In Figure \ref{co_full}, we have shown the positions of detected CO clouds (green circles) in the two targeted fields (north-west (NW) and south-east (SE)) shown in black square boxes. The same blue and red contours of Figure \ref{caf2_b4} are also shown to correlate the position of detected molecular clouds with respect to the hot star forming regions. In Figure \ref{co_zoom}, we zoom in to these regions where CO clouds are detected (black box of Figure \ref{co_full}). Along with the spatial position, we show the size and virial mass of each cloud as estimated by \citet{rubio2015}. The majority of the detected clouds are of low mass ($\sim$ 10$^3$ $M_{\odot}$), which in principle supports our finding of a large number of low mass compact star forming regions in WLM. Again the absence of higher mass molecular clouds could be the reason for the absence of massive compact star forming regions in this galaxy. We further find that eight of the detected CO clouds are found to be present away from the hot star forming regions. The only two clouds which are found to be co-located (there could also be a projection effect) with the hot star forming regions are NW-1 and SE-4, where SE-4 is the most massive cloud detected.

 \subsection{Star Formation Rate}
\begin{figure}
\begin{center}
\includegraphics[width=3.7in]{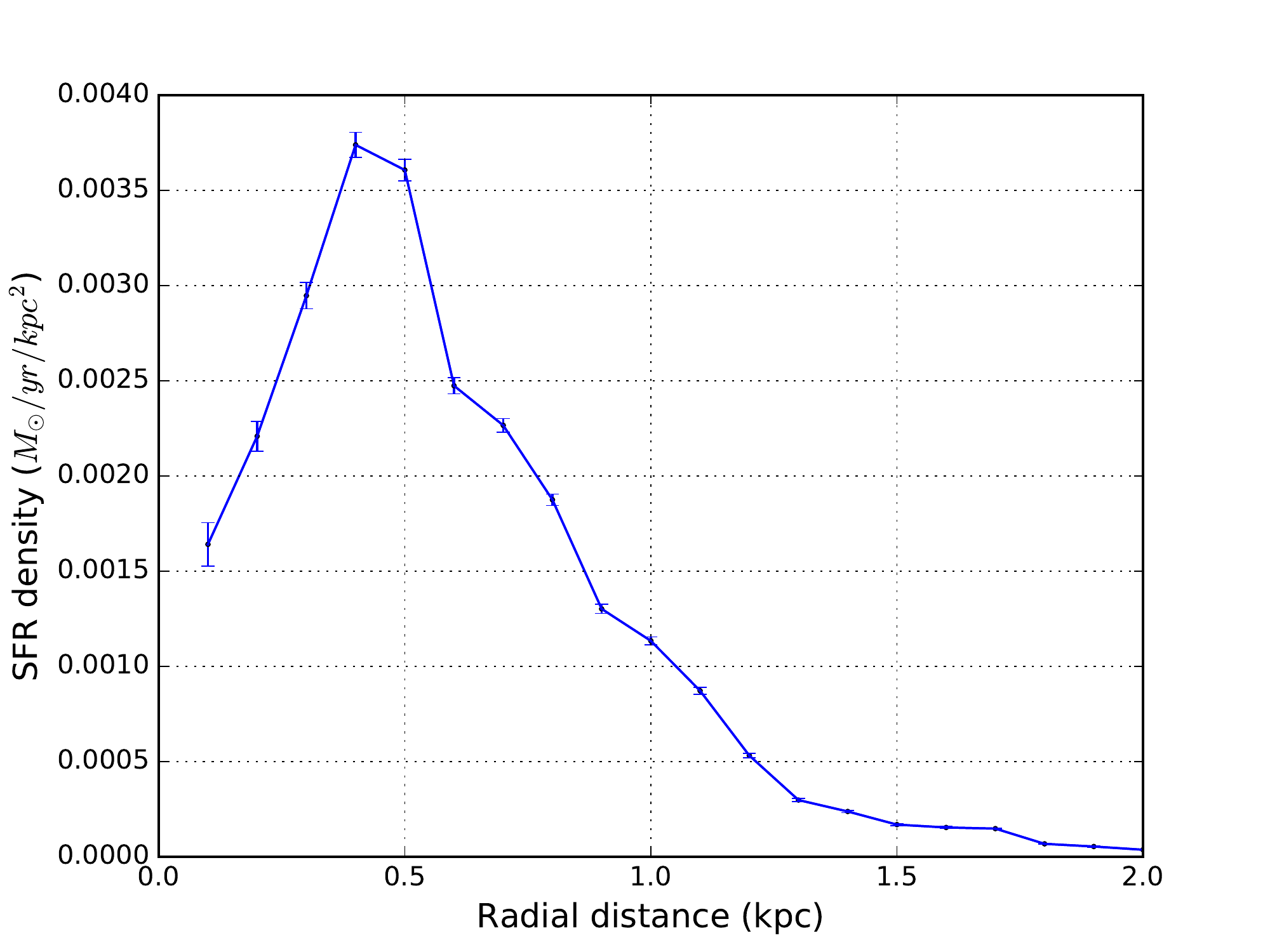} 
\caption{The radial profile for SFR density ($M_{\odot}/yr/kpc^2$) with 3$\sigma$ error bar is shown in the figure.}
\label{sfr_radial}
\end{center}
\end{figure}

\begin{figure}
\begin{center}
\includegraphics[width=2.7in]{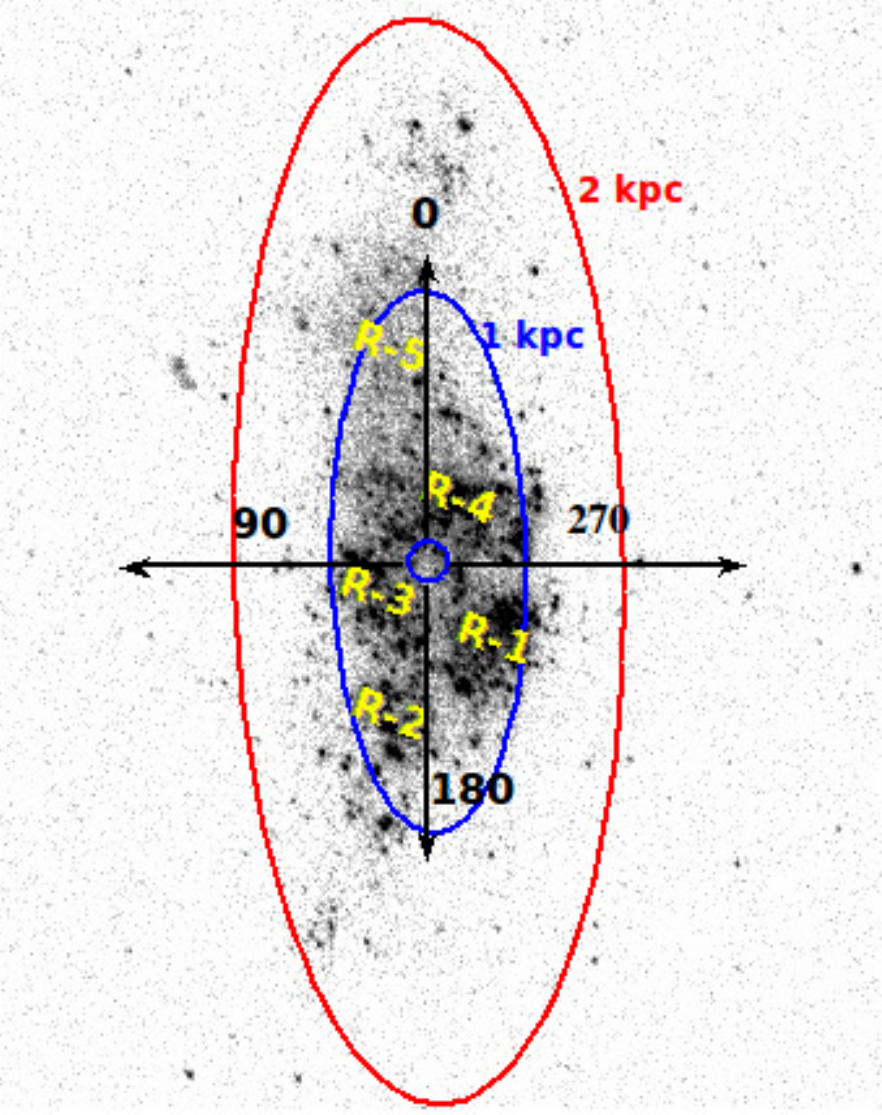} 
 \caption{The figure shows all five regions (R1, R2, R3, R4 and R5) with F148W band image in the background. The black arrows show the position angle along four different directions. The blue and red ellipses represent 1 kpc and 2 kpc galactocentric distance respectively. }
 \label{pa}
 \end{center}
 \end{figure}

\begin{figure}
\begin{center}
\includegraphics[width=3.7in]{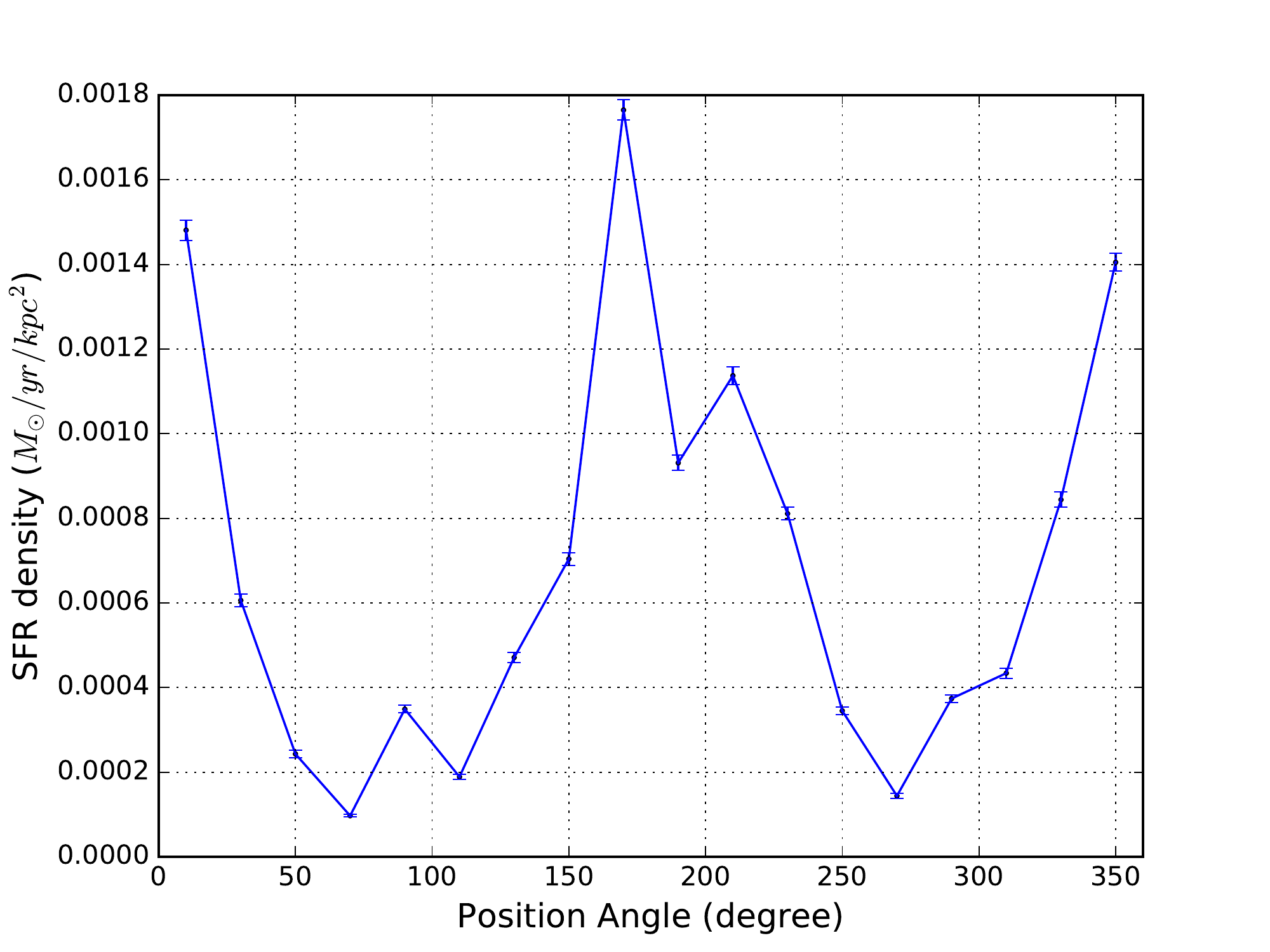} 
 \caption{The azimuthal variation of radially averaged (up to 2 kpc) SFR density ($M_{\odot}/yr/kpc^2$) with 3$\sigma$ error bar is shown in the figure. Two peaks at PA $\sim$ 0$^o$ and 180$^o$ signify the contribution of star forming regions from northern and southern half of the galaxy respectively.} 
 \label{sfr_azimuth}
 \end{center}
 \end{figure}

 In order to estimate the star formation rate (SFR) of WLM, we used the F148W band (FUV) exposure normalized image of the galaxy. We summed up the flux (CPS) within a galactocentric distance of 1.7 kpc, which is the extent of WLM's outer disk as found in section \ref{lum_s}.
 The background corrected total flux is converted to magnitude and further corrected for extinction.  The background and extinction corrected magnitude is used to calculate the total SFR of the galaxy ($M_{\odot}/yr$). The corrected FUV flux is found to be 7.0$\times10^{-13}$ $\pm$ 3.6$\times10^{-15} erg/sec/cm^2/\AA$ and the corresponding magnitude to be 12.12$\pm$0.01. We used the scaling relation, given in equation \ref{sfr_eq} \citep{kara2013}, to estimate the SFR, where $mag_{FUV}$ is the corrected FUV magnitude and $D$ is the distance to the galaxy in Mpc. \citet{kara2013} used this relation to estimate the SFR of a large sample of galaxies in the local volume using GALEX FUV observation. Since F148W filter of UVIT has bandpass similar to GALEX FUV, we used this relation to estimate SFR. The total SFR up to a radial distance 1.7 kpc is found to be 0.008$\pm$0.0001 $M_{\odot}/yr$. This value of SFR  could change if we adopt a different scaling relation. For example, if we adopt the empirical relation modelled by \citet{mcquinn2015}, we estimate the SFR of WLM in the range 0.002 - 0.006 $M_{\odot}/yr$. \\\\
In order to explore radial variation of SFR within WLM, we generated the radial profile of SFR density (i.e. $M_{\odot}/yr/kpc^2$).
The radial variation of SFR density is shown in Figure \ref{sfr_radial}. The SFR density is found to increase with radius up to 0.4 kpc, resulting in a peak at this radius. The SFR density decreases after 0.4 kpc in the outer part, where the profile is shown up to 2 kpc. The profile also suggests that the SFR density is dominated by regions within a radius of 1 kpc. The relative positions of all the regions (R1, R2, R3, R4) along with the position angle is shown in Figure \ref{pa}. This figure shows that most of these regions are located within the 1 kpc radius. We further explored the azimuthal variation of SFR density by taking the radially averaged value up to 2 kpc in Figure \ref{sfr_azimuth}.  The profile shows two peaks at diametrically opposite values of position angle ($\sim$ 0$^o$ and 180$^o$). The peak near PA $\sim$ 0$^o$ has contribution from the northern regions R4 and R5, whereas the peak at PA $\sim$ 180$^o$ has contributions from R1, R2 and R3.  Figure \ref{sfr_azimuth} and \ref{pa} suggest that the southern half of the galaxy has a slightly higher SFR density when compared to the northern half.  
 
 \begin{equation}
log(SFR_{FUV} (M_{\odot}/yr)) = 2.78 - 0.4*mag_{FUV} + 2log(D) 
\label{sfr_eq}
 \end{equation}
 
\section{Discussion}
The primary objective of this study is to identify hot star forming regions of the dwarf irregular galaxy WLM and map its demographics. We used the UV images of the galaxy acquired from UVIT having a superior spatial resolution ($\sim$ 1.5 arcsec).  The hot star forming regions are identified from the UV color maps with the help of temperature - flux ratio relation derived from Kurucz spectra.  We point out that this relation may change for any corresponding change in the assumed value of metallicity, extinction and surface gravity of stars.
 We  demonstrate that the change of metallicity value between 0.002 to 0.006 has negligible effect on the flux ratio, hence on the measured temperature. On the other hand, the estimated parameters will have a dependence on the assumed extinction and its variation. Figure \ref{kuruz_temp}a shows the range of temperature for a given flux ratio, for various extinction laws. It also suggests that the uncertainty in temperature due to extinction is more for a higher temperature. \\\\ 

For any dwarf galaxy, feedback from massive stars act as one of the dominant mechanisms for driving star formation. 
According to the study of \citet{dolphin2000}, WLM formed half of its stars, 9 Gyr ago and shows a reduction in star formation rate after that. They also found the star forming activity to increase around 1 to 2.5 Gyr, which continued till now.  \citet{bianchi2012} found the galaxy to undergo a vigorous star formation during the last 10 Myr. They found 531 stars with temperature T $>$ 18000 K and 721 stars having M $>$ 9 $M_{\odot}$.  In this study, we detect several hot regions with T $>$ 35000 K, which are likely to be populated by O type stars. As the life time of these stars are  $\sim$ 10 Myr, this would suggest that WLM has undergone a recent star formation activity with the production of massive stars.  We also notice the hot regions to be surrounded by relatively cooler regions, suggestive of hierarchical structure. These are in agreement with the findings of \citet{bianchi2012}. We identified around 20 hot complexes with a size of 20 - 50 pc and more than 50 complexes with a size of 10 - 20 pc. Regions with  size $<$ 10 pc  are found to be large in number (upto a few hundred) and scattered over a larger area of the galaxy. The smaller regions have less masses and are likely to be populated by a few OB stars, whereas the larger ones are likely to be complexes consisting of two or more associations of the galaxy. Our study finds that WLM is dominated by small sized regions, with the largest complex not exceeding 50 pc.\\

It is well known that the star formation creates hierarchical structures comprising of star clusters, OB associations, and complexes \citep{efremov1998,bastian2007,sun2018}. These structures are found to have various sizes and density in various environments \citep{elmegreen2014,grasha2017}. \citet{bresolin1998} studied OB associations in 7 nearby spiral galaxies using HST-WFPC2 data and found a size range of $\sim$ 70 - 100 pc. For nearby irregular galaxies, the size of OB associations are reported to be $\sim$ 80 pc by \citet{ivanov1996}. \citet{melnik1995} found relatively smaller size ($\sim$ 10 - 80 pc, with a peak around 30 pc) for galactic OB association whereas for LMC and SMC, it is around 60 pc \citep{bresolin1996}. \citet{bastian2007} noticed the star formation in M33 to be hierarchical in nature with a size distribution of young stellar complexes between 10 - 100 pc. \citet{garcia2010} studied IC 1613, a similar nearby dwarf irregular galaxy like WLM, and concluded the star formation to be hierarchical with no specific length scale.  The identified OB associations in their study show a size $\sim$ 14 - 42 pc which is quite similar to our findings in WLM.  \citet{garcia2010} also found 44\% of the total associations are smaller in size. Several studies conclude that the size of these detected associations is highly dependent on the resolution of telescope and the distance to the galaxy \citep{garcia2010,bastian2007}. Due to the better spatial resolution of UVIT we are able to detect smaller substructures of the star forming regions present in the galaxy, when compared to GALEX. The overall sizes of OB associations in WLM, detected through UVIT are found to be quite similar to those in IC 1613 (nearby dwarf irregular), but smaller than those in M33 (nearby spiral) and the Milky Way. \\

Around most of the larger hot complexes we detect the presence of several smaller regions connected together by low temperature contours. This brings out the clumpy nature of star forming regions in WLM.  The feedback from the massive stars (large hot complex) may have triggered further star formation near to them which results in these clumpy distribution of young star forming regions. It is also noticed that relatively low temperature regions are much more spread than the hot regions which is also mentioned in the study by \citet{bianchi2012}. The star forming regions also show several knotted structures, specifically in R1. The overall distribution of detected hot star forming regions throughout the galaxy clearly reveals that the southern part of the galaxy is much more active than its northern part. The hot star forming regions detected in (F148W$-$N245M) color map are found to be more fragmented than those seen in (F148W$-$N263M) color map. A clumpy distribution of either the young stellar objects or the dust or both of them present in the galaxy can result in these observed structural differences.

We also studied the extent of UV emission in WLM. It is noticed that the NUV emission as traced by N245M and N263M filters is more extended than the FUV emission. Since the FUV emission mainly comes from the younger stellar populations, it is probable that the populations present in the outer part of the galaxy are relatively older than those of inner part. The luminosity density profile clearly shows that the UV emission of the galaxy is present at least up to a radius 1.7 kpc.\\\\ 
A high resolution H$~$I study by \citet{kepley2007} revealed the presence of a hook like structure in H$~$I distribution. They speculated that an expanding ring like structure has broken in the western part and generated this hook like pattern. They concluded that this structure is likely to be the result of star formation propagating outside from the center of the galaxy. The hot regions detected in our study delineate this hook structure except in the western part (i.e. in R1). The H$~$I density shows an overall reduction in R1 whereas we see intense H$\alpha$ emission with a central hole in the same region. On the basis of this scenario \citet{kepley2007} concluded that H$~$I is likely to be used up or ionized or destroyed due to high rate of star formation in this part of the galaxy. In our study we detected several knotted structures of hot star forming regions in R1 which signify the presence of several OB stars. This region also has one of the three most luminous star forming complexes (see Figure \ref{caf2_mass}). This L-shaped complex is located to the north and is suggestive of intense recent star formation. We also notice the presence of hot star forming regions forming an arc like shape, located to the west of the H-alpha ring (see Figure \ref{Halpha}). This might suggest that the H-alpha emission is shrinking towards the inner hole, in the eastern side. The western arc like structure could be relatively older with respect to the northern complex. The radiation from these OB stars has probably driven/consumed the H$~$I gas present there, which supports the scenario of \citet{kepley2007}.\\\\
In a recent study with ALMA observations, \citet{rubio2015} detected several CO clouds with sizes ranging from 3 to 12 pc in diameter. The smaller hot star forming regions (i.e. clusters or OB associations) identified in our study have an average size from 4 - 20 pc, with most of them with sizes $<$ 10 pc. The comparable size of CO clouds and hot star forming regions suggest that star formation happens in smaller molecular clouds in this galaxy. The masses estimated for the compact star forming regions detected in this study also match well with the mass of the CO clouds, confirming the formation of low mass stellar clumps.\\\\
We estimated the total FUV magnitude to be 12.12$\pm$0.01 magnitude, corresponding to a SFR of $\sim$0.008$\pm$0.0001 $M_{\odot}/yr$. As mentioned earlier, the value of SFR is a function of the scaling relation and could change if a different relation is used. Earlier estimations by \citet{kara2013} and \citet{hunter2010} using GALEX FUV observations are comparable. Though SFR of WLM is less compared to that of Milky way (1.9$\pm$0.4 $M_{\odot}/yr$ \citep{chomiuk2011}), the value of specific SFR (sSFR = SFR/Stellar mass of galaxy) of WLM is 16 times higher than that of Milky Way. For the total stellar mass of WLM ($1.6\times10^7 M_{\odot}$ \citep{zhang2012}) and Milky Way ($(6.4\pm0.6)\times10^{10} M_{\odot}$ \citep{mcmillan2011}), the estimated values of sSFR are $5.0\times10^{-10}$ and $3.0\times10^{-11}$ $yr^{-1}$ respectively. Therefore being a dwarf irregular galaxy, WLM is able to form stars in a relatively higher rate than Milky way \citep{rubio2015}.

\section{Summary}
The main results of this study are summarized below:
\begin{enumerate}
\item The UVIT images in FUV and NUV are used to study the demographics of the star forming regions in WLM.
\item The UV emission of WLM is found to be present up to a galactocentric distance of 1.7 kpc, with most of the flux originating within a radius of 1 kpc. 
 \item We divided the galaxy into 5 different regions to compare the demographics and identify three most luminous and massive star forming complexes, with the largest one extending not more than 50pc.
  \item The hot star forming regions with T $>$ 17500 K (populated with OB stars) have a range in size, with a few tens of regions in the 20-50 pc range and a few hundred regions with $<$ 10 pc size. 
  \item The hot stellar regions are found to be surrounded by regions with cooler stars, suggesting a hierarchical nature of distribution. 
  \item The presence of several hot regions suggest that the galaxy underwent a recent star formation  $\sim$ 10 Myr ago. 
  \item We find a good spatial correlation between UV detected hot regions with H$\alpha$ emitting regions, H$~$I distribution and HST detected massive stars in the galaxy. The western part (R1) of the galaxy is likely to have undergone a vigorous recent star formation.
  \item The sizes and masses of isolated star forming regions closely match with those of CO clouds estimated using ALMA observations. WLM is found to have a large number of low mass (M $< 10^3 M_{\odot}$) compact star forming regions.
  \item The star formation in the southern part of the galaxy is more active than the northern part. The SFR of the galaxy is $\sim$ 0.008 $M_{\odot}/yr$, which is similar to that
  found in other dwarf galaxies.
\end{enumerate}

\acknowledgments
UVIT project is a result of collaboration between IIA, Bengaluru, IUCAA, Pune, TIFR, Mumbai, several centres of ISRO, and CSA. Indian Institutions and the Canadian Space Agency have contributed to the work presented in this paper. Several groups from ISAC (ISRO), Bengaluru, and IISU (ISRO), Trivandrum have contributed to the design, fabrication, and testing of the payload. The Mission Group (ISAC) and ISTRAC (ISAC) continue to provide support in making observations with, and reception and initial processing of the data.  We gratefully thank all the individuals involved in the various teams for providing their support to the project from the early stages of the design to launch and observations with it in the orbit. Finally, we thank the referee for valuable suggestions.


\begin{thebibliography}{}
\expandafter\ifx\csname natexlab\endcsname\relax\def\natexlab#1{#1}\fi
\providecommand{\url}[1]{\href{#1}{#1}}

\bibitem[{{Bastian} {et~al.}(2007){Bastian}, {Ercolano}, {Gieles},
  {Rosolowsky}, {Scheepmaker}, {Gutermuth}, \& {Efremov}}]{bastian2007}
{Bastian}, N., {Ercolano}, B., {Gieles}, M., {et~al.} 2007, \mnras, 379, 1302

\bibitem[{{Bianchi} {et~al.}(2012){Bianchi}, {Efremova}, {Hodge}, {Massey}, \&
  {Olsen}}]{bianchi2012}
{Bianchi}, L., {Efremova}, B., {Hodge}, P., {Massey}, P., \& {Olsen}, K.~A.~G.
  2012, \aj, 143, 74

\bibitem[{{Bresolin} {et~al.}(1996){Bresolin}, {Kennicutt}, \&
  {Stetson}}]{bresolin1996}
{Bresolin}, F., {Kennicutt}, Jr., R.~C., \& {Stetson}, P.~B. 1996, \aj, 112,
  1009

\bibitem[{{Bresolin} {et~al.}(1998){Bresolin}, {Kennicutt}, {Ferrarese},
  {Gibson}, {Graham}, {Macri}, {Phelps}, {Rawson}, {Sakai}, {Silbermann},
  {Stetson}, \& {Turner}}]{bresolin1998}
{Bresolin}, F., {Kennicutt}, Jr., R.~C., {Ferrarese}, L., {et~al.} 1998, \aj,
  116, 119

\bibitem[{{Calzetti} {et~al.}(1994){Calzetti}, {Kinney}, \&
  {Storchi-Bergmann}}]{calzetti1994}
{Calzetti}, D., {Kinney}, A.~L., \& {Storchi-Bergmann}, T. 1994, \apj, 429, 582

\bibitem[{{Castelli} \& {Kurucz}(2004)}]{castelli2004}
{Castelli}, F., \& {Kurucz}, R.~L. 2004, ArXiv Astrophysics e-prints,
  astro-ph/0405087

\bibitem[{{Chomiuk} \& {Povich}(2011)}]{chomiuk2011}
{Chomiuk}, L., \& {Povich}, M.~S. 2011, \aj, 142, 197

\bibitem[{{Cignoni} {et~al.}(2018){Cignoni}, {Sacchi}, {Aloisi}, {Tosi},
  {Calzetti}, {Lee}, {Sabbi}, {Adamo}, {Cook}, {Dale}, {Elmegreen},
  {Gallagher}, {Gouliermis}, {Grasha}, {Grebel}, {Hunter}, {Johnson}, {Messa},
  {Smith}, {Thilker}, {Ubeda}, \& {Whitmore}}]{cignoni2018}
{Cignoni}, M., {Sacchi}, E., {Aloisi}, A., {et~al.} 2018, \apj, 856, 62

\bibitem[{{Cole} {et~al.}(2007){Cole}, {Skillman}, {Tolstoy}, {Gallagher},
  {Aparicio}, {Dolphin}, {Gallart}, {Hidalgo}, {Saha}, {Stetson}, \&
  {Weisz}}]{cole2007}
{Cole}, A.~A., {Skillman}, E.~D., {Tolstoy}, E., {et~al.} 2007, \apjl, 659, L17

\bibitem[{{de Vaucouleurs}(1991)}]{devaucouleurs1991}
{de Vaucouleurs}, G. 1991, Science, 254, 1667

\bibitem[{{Dolphin}(2000)}]{dolphin2000}
{Dolphin}, A.~E. 2000, \apj, 531, 804

\bibitem[{{Efremov} \& {Elmegreen}(1998)}]{efremov1998}
{Efremov}, Y.~N., \& {Elmegreen}, B.~G. 1998, \mnras, 299, 588

\bibitem[{{Elmegreen} {et~al.}(2014){Elmegreen}, {Elmegreen}, {Adamo},
  {Aloisi}, {Andrews}, {Annibali}, {Bright}, {Calzetti}, {Cignoni}, {Evans},
  {Gallagher}, {Gouliermis}, {Grebel}, {Hunter}, {Johnson}, {Kim}, {Lee},
  {Sabbi}, {Smith}, {Thilker}, {Tosi}, \& {Ubeda}}]{elmegreen2014}
{Elmegreen}, D.~M., {Elmegreen}, B.~G., {Adamo}, A., {et~al.} 2014, \apjl, 787,
  L15

\bibitem[{{Ferraro} {et~al.}(1989){Ferraro}, {Fusi Pecci}, {Tosi}, \&
  {Buonanno}}]{ferraro1989}
{Ferraro}, F.~R., {Fusi Pecci}, F., {Tosi}, M., \& {Buonanno}, R. 1989, \mnras,
  241, 433

\bibitem[{{Fitzpatrick}(1999)}]{fitzpatrick1999}
{Fitzpatrick}, E.~L. 1999, \pasp, 111, 63

\bibitem[{{Gallouet} {et~al.}(1975){Gallouet}, {Heidmann}, \&
  {Dampierre}}]{gallouet1975}
{Gallouet}, L., {Heidmann}, N., \& {Dampierre}, F. 1975, \aaps, 19, 1

\bibitem[{{Garcia} {et~al.}(2010){Garcia}, {Herrero}, {Castro}, {Corral}, \&
  {Rosenberg}}]{garcia2010}
{Garcia}, M., {Herrero}, A., {Castro}, N., {Corral}, L., \& {Rosenberg}, A.
  2010, \aap, 523, A23

\bibitem[{{George} {et~al.}(2018{\natexlab{a}}){George}, {Joseph},
  {C{\^o}t{\'e}}, {Ghosh}, {Hutchings}, {Mohan}, {Postma},
  {Sankarasubramanian}, {Sreekumar}, {Stalin}, {Subramaniam}, \&
  {Tandon}}]{koshy2018a}
{George}, K., {Joseph}, P., {C{\^o}t{\'e}}, P., {et~al.} 2018{\natexlab{a}},
  ArXiv e-prints, arXiv:1802.09493

\bibitem[{{George} {et~al.}(2018{\natexlab{b}}){George}, {Poggianti},
  {Gullieuszik}, {Fasano}, {Bellhouse}, {Postma}, {Moretti}, {Jaff{\'e}},
  {Vulcani}, {Bettoni}, {Fritz}, {C{\^o}t{\'e}}, {Ghosh}, {Hutchings}, {Mohan},
  {Sreekumar}, {Stalin}, {Subramaniam}, \& {Tandon}}]{koshy2018b}
{George}, K., {Poggianti}, B.~M., {Gullieuszik}, M., {et~al.}
  2018{\natexlab{b}}, ArXiv e-prints, arXiv:1803.06193

\bibitem[{{George} {et~al.}(2018{\natexlab{c}}){George}, {Joseph}, {Mondal},
  {Devaraj}, {Subramaniam}, {Stalin}, {C{\^o}t{\'e}}, {Ghosh}, {Hutchings},
  {Mohan}, {Postma}, {Sankarasubramanian}, {Sreekumar}, \&
  {Tandon}}]{koshy2018c}
{George}, K., {Joseph}, P., {Mondal}, C., {et~al.} 2018{\natexlab{c}}, ArXiv
  e-prints, arXiv:1805.03543

\bibitem[{{Gieren} {et~al.}(2008){Gieren}, {Pietrzy{\'n}ski}, {Szewczyk},
  {Soszy{\'n}ski}, {Bresolin}, {Kudritzki}, {Urbaneja}, {Storm}, \&
  {Minniti}}]{gieren2008}
{Gieren}, W., {Pietrzy{\'n}ski}, G., {Szewczyk}, O., {et~al.} 2008, \apj, 683,
  611

\bibitem[{{Gil de Paz} {et~al.}(2007){Gil de Paz}, {Boissier}, {Madore},
  {Seibert}, {Joe}, {Boselli}, {Wyder}, {Thilker}, {Bianchi}, {Rey}, {Rich},
  {Barlow}, {Conrow}, {Forster}, {Friedman}, {Martin}, {Morrissey}, {Neff},
  {Schiminovich}, {Small}, {Donas}, {Heckman}, {Lee}, {Milliard}, {Szalay}, \&
  {Yi}}]{gildepaz2007}
{Gil de Paz}, A., {Boissier}, S., {Madore}, B.~F., {et~al.} 2007, \apjs, 173,
  185

\bibitem[{{Girish} {et~al.}(2017){Girish}, {Tandon}, {Sriram}, {Kumar}, \&
  {Postma}}]{girish2017}
{Girish}, V., {Tandon}, S.~N., {Sriram}, S., {Kumar}, A., \& {Postma}, J. 2017,
  Experimental Astronomy, 43, 59

\bibitem[{{Gordon} {et~al.}(2003){Gordon}, {Clayton}, {Misselt}, {Landolt}, \&
  {Wolff}}]{gordon2003}
{Gordon}, K.~D., {Clayton}, G.~C., {Misselt}, K.~A., {Landolt}, A.~U., \&
  {Wolff}, M.~J. 2003, \apj, 594, 279

\bibitem[{{Grasha} {et~al.}(2017){Grasha}, {Calzetti}, {Adamo}, {Kim},
  {Elmegreen}, {Gouliermis}, {Dale}, {Fumagalli}, {Grebel}, {Johnson}, {Kahre},
  {Kennicutt}, {Messa}, {Pellerin}, {Ryon}, {Smith}, {Shabani}, {Thilker}, \&
  {Ubeda}}]{grasha2017}
{Grasha}, K., {Calzetti}, D., {Adamo}, A., {et~al.} 2017, \apj, 840, 113

\bibitem[{{Hunter}(1997)}]{hunter1997}
{Hunter}, D. 1997, \pasp, 109, 937

\bibitem[{{Hunter} {et~al.}(2010){Hunter}, {Elmegreen}, \&
  {Ludka}}]{hunter2010}
{Hunter}, D.~A., {Elmegreen}, B.~G., \& {Ludka}, B.~C. 2010, \aj, 139, 447

\bibitem[{{Ivanov}(1996)}]{ivanov1996}
{Ivanov}, G.~R. 1996, \aap, 305, 708

\bibitem[{{Karachentsev} \& {Kaisina}(2013)}]{kara2013}
{Karachentsev}, I.~D., \& {Kaisina}, E.~I. 2013, \aj, 146, 46

\bibitem[{{Kepley} {et~al.}(2007){Kepley}, {Wilcots}, {Hunter}, \&
  {Nordgren}}]{kepley2007}
{Kepley}, A.~A., {Wilcots}, E.~M., {Hunter}, D.~A., \& {Nordgren}, T. 2007,
  \aj, 133, 2242

\bibitem[{{Kroupa}(2001)}]{kroupa2001}
{Kroupa}, P. 2001, \mnras, 322, 231

\bibitem[{{Kumar} {et~al.}(2012){Kumar}, {Ghosh}, {Hutchings}, {Kamath},
  {Kathiravan}, {Mahesh}, {Murthy}, {Nagbhushana}, {Pati}, {Rao}, {Rao},
  {Sriram}, \& {Tandon}}]{kumar2012}
{Kumar}, A., {Ghosh}, S.~K., {Hutchings}, J., {et~al.} 2012, in \procspie, Vol.
  8443, Space Telescopes and Instrumentation 2012: Ultraviolet to Gamma Ray,
  84431N

\bibitem[{{Leaman} {et~al.}(2012){Leaman}, {Venn}, {Brooks}, {Battaglia},
  {Cole}, {Ibata}, {Irwin}, {McConnachie}, {Mendel}, \& {Tolstoy}}]{leaman2012}
{Leaman}, R., {Venn}, K.~A., {Brooks}, A.~M., {et~al.} 2012, \apj, 750, 33

\bibitem[{{Leitherer} {et~al.}(1999){Leitherer}, {Schaerer}, {Goldader},
  {Delgado}, {Robert}, {Kune}, {de Mello}, {Devost}, \&
  {Heckman}}]{leitherer1999}
{Leitherer}, C., {Schaerer}, D., {Goldader}, J.~D., {et~al.} 1999, \apjs, 123,
  3

\bibitem[{{Massey} {et~al.}(2007){Massey}, {McNeill}, {Olsen}, {Hodge},
  {Blaha}, {Jacoby}, {Smith}, \& {Strong}}]{massey2007}
{Massey}, P., {McNeill}, R.~T., {Olsen}, K.~A.~G., {et~al.} 2007, \aj, 134,
  2474

\bibitem[{{McCall}(2004)}]{mccall2004}
{McCall}, M.~L. 2004, \aj, 128, 2144

\bibitem[{{McMillan}(2011)}]{mcmillan2011}
{McMillan}, P.~J. 2011, \mnras, 414, 2446

\bibitem[{{McQuinn} {et~al.}(2015){McQuinn}, {Skillman}, {Dolphin}, \&
  {Mitchell}}]{mcquinn2015}
{McQuinn}, K.~B.~W., {Skillman}, E.~D., {Dolphin}, A.~E., \& {Mitchell}, N.~P.
  2015, \apj, 808, 109

\bibitem[{{Melena} {et~al.}(2009){Melena}, {Elmegreen}, {Hunter}, \&
  {Zernow}}]{melena2009}
{Melena}, N.~W., {Elmegreen}, B.~G., {Hunter}, D.~A., \& {Zernow}, L. 2009,
  \aj, 138, 1203

\bibitem[{{Mel'Nik} \& {Efremov}(1995)}]{melnik1995}
{Mel'Nik}, A.~M., \& {Efremov}, Y.~N. 1995, Astronomy Letters, 21, 10

\bibitem[{{Minniti} \& {Zijlstra}(1997)}]{minniti1997}
{Minniti}, D., \& {Zijlstra}, A.~A. 1997, \aj, 114, 147

\bibitem[{{Postma} {et~al.}(2011){Postma}, {Hutchings}, \&
  {Leahy}}]{postma2011}
{Postma}, J., {Hutchings}, J.~B., \& {Leahy}, D. 2011, \pasp, 123, 833

\bibitem[{{Postma} \& {Leahy}(2017)}]{postma2017}
{Postma}, J.~E., \& {Leahy}, D. 2017, \pasp, 129, 115002

\bibitem[{{Rejkuba} {et~al.}(2000){Rejkuba}, {Minniti}, {Gregg}, {Zijlstra},
  {Alonso}, \& {Goudfrooij}}]{rejkuba2000}
{Rejkuba}, M., {Minniti}, D., {Gregg}, M.~D., {et~al.} 2000, \aj, 120, 801

\bibitem[{{Rubio} {et~al.}(2015){Rubio}, {Elmegreen}, {Hunter}, {Brinks},
  {Cort{\'e}s}, \& {Cigan}}]{rubio2015}
{Rubio}, M., {Elmegreen}, B.~G., {Hunter}, D.~A., {et~al.} 2015, \nat, 525, 218

\bibitem[{{Subramaniam} {et~al.}(2017){Subramaniam}, {Sahu}, {Postma},
  {C{\^o}t{\'e}}, {Hutchings}, {Darukhanawalla}, {Chung}, {Tandon}, {Kameswara
  Rao}, {George}, {Ghosh}, {Girish}, {Mohan}, {Murthy}, {Pati},
  {Sankarasubramanian}, {Stalin}, \& {Choudhury}}]{subramaniam2017}
{Subramaniam}, A., {Sahu}, S., {Postma}, J.~E., {et~al.} 2017, \aj, 154, 233

\bibitem[{{Sun} {et~al.}(2018){Sun}, {de Grijs}, {Cioni}, {Rubele},
  {Subramanian}, {van Loon}, {Bekki}, {Bell}, {Ivanov}, {Marconi}, {Muraveva},
  {Oliveira}, \& {Ripepi}}]{sun2018}
{Sun}, N.-C., {de Grijs}, R., {Cioni}, M.-R.~L., {et~al.} 2018, \apj, 858, 31

\bibitem[{{Tandon} {et~al.}(2017){Tandon}, {Subramaniam}, {Girish}, {Postma},
  {Sankarasubramanian}, {Sriram}, {Stalin}, {Mondal}, {Sahu}, {Joseph},
  {Hutchings}, {Ghosh}, {Barve}, {George}, {Kamath}, {Kathiravan}, {Kumar},
  {Lancelot}, {Leahy}, {Mahesh}, {Mohan}, {Nagabhushana}, {Pati}, {Kameswara
  Rao}, {Sreedhar}, \& {Sreekumar}}]{tandon2017}
{Tandon}, S.~N., {Subramaniam}, A., {Girish}, V., {et~al.} 2017, \aj, 154, 128

\bibitem[{{Thilker} {et~al.}(2007){Thilker}, {Bianchi}, {Meurer}, {Gil de Paz},
  {Boissier}, {Madore}, {Boselli}, {Ferguson}, {Mu{\~n}oz-Mateos}, {Madsen},
  {Hameed}, {Overzier}, {Forster}, {Friedman}, {Martin}, {Morrissey}, {Neff},
  {Schiminovich}, {Seibert}, {Small}, {Wyder}, {Donas}, {Heckman}, {Lee},
  {Milliard}, {Rich}, {Szalay}, {Welsh}, \& {Yi}}]{thilker2007}
{Thilker}, D.~A., {Bianchi}, L., {Meurer}, G., {et~al.} 2007, \apjs, 173, 538

\bibitem[{{Tolstoy} {et~al.}(2009){Tolstoy}, {Hill}, \& {Tosi}}]{tolstoy2009}
{Tolstoy}, E., {Hill}, V., \& {Tosi}, M. 2009, \araa, 47, 371

\bibitem[{{Urbaneja} {et~al.}(2008){Urbaneja}, {Kudritzki}, {Bresolin},
  {Przybilla}, {Gieren}, \& {Pietrzy{\'n}ski}}]{urbaneja2008}
{Urbaneja}, M.~A., {Kudritzki}, R.-P., {Bresolin}, F., {et~al.} 2008, \apj,
  684, 118

\bibitem[{{van der Marel} \& {Cioni}(2001)}]{marel2001}
{van der Marel}, R.~P., \& {Cioni}, M.-R.~L. 2001, \aj, 122, 1807

\bibitem[{{Weisz} {et~al.}(2011){Weisz}, {Dalcanton}, {Williams}, {Gilbert},
  {Skillman}, {Seth}, {Dolphin}, {McQuinn}, {Gogarten}, {Holtzman}, {Rosema},
  {Cole}, {Karachentsev}, \& {Zaritsky}}]{weisz2011}
{Weisz}, D.~R., {Dalcanton}, J.~J., {Williams}, B.~F., {et~al.} 2011, \apj,
  739, 5

\bibitem[{{Whiting} {et~al.}(1999){Whiting}, {Hau}, \& {Irwin}}]{whiting1999}
{Whiting}, A.~B., {Hau}, G.~K.~T., \& {Irwin}, M. 1999, \aj, 118, 2767

\bibitem[{{Zhang} {et~al.}(2012){Zhang}, {Hunter}, {Elmegreen}, {Gao}, \&
  {Schruba}}]{zhang2012}
{Zhang}, H.-X., {Hunter}, D.~A., {Elmegreen}, B.~G., {Gao}, Y., \& {Schruba},
  A. 2012, \aj, 143, 47

\end{thebibliography}

\end{document}